\documentclass[final,3p,times,twocolumn]{elsarticle}

\usepackage{amssymb}
\usepackage{amsmath}

 \biboptions{comma, square}

\newcommand{\met}{\mbox{$\protect \raisebox{0.3ex}{{$\not$}E}_T$}}
\newcommand{\mpt}{\mbox{$\protect \raisebox{0.3ex}{{$\not$}p}_T$}}

\newcommand{\vecmet}{\mbox{$\protect \raisebox{0.3ex}{$\not$}\vec{E}_T$}}
\newcommand{\vecmpt}{\mbox{$\protect \raisebox{0.3ex}{$\not$}\vec{p}_T$}}

\journal{Nuclear Physics A}

\begin{document}

\begin{frontmatter}

\title{Detection of invisible particles at hadron collider experiments through the magnetic spectrometer}

 \author[label1]{Marco Bentivegna}
 \address[label1]{Sapienza Universit\`a di Roma, I-00185 Roma, Italy}
 \author[label2]{Qiuguang Liu}
 \address[label2]{Purdue University, West Lafayette, Indiana 47907, USA}
 \author[label1]{Fabrizio Margaroli}
 \author[label2]{Karolos Potamianos}

\begin{abstract}

The production of invisible particles plays great importance in high energy physics. Large part of interesting electroweak processes include production of neutrinos, while many new physics scenarios predict the existence of similarly weakly-interacting particles. In events with associated production of invisible particles and hadronic jets, the measurement of the imbalance in transverse momentum of the final state particles is the major leverage to reject the otherwise dominant source of backgrounds in hadron colliders,\, {\it i.e.}\,the generic production of many jets by QCD interactions. Here we discuss a novel technique which utilizes the information derived from the spectrometer, eventually coupled with the more straightforward calorimeter information, to infer the passage of invisible particles. We check the validity of this technique in data and Monte Carlo simulations in a broad range of topologies, starting from the simplest, with two jets in the final state, to the ones with very large jet multiplicities. We also suggest a new way, based on the same approach, to measure the yields and model the kinematics of the QCD multijet background in invisible particles plus jets signatures. The results are derived using data collected with the CDF\,II detector; we argue that the application to  LHC experiments is straightforward.
\end{abstract}

\begin{keyword}

Collider physics  \sep QCD multijet \sep background rejection \sep missing transverse energy \sep missing transverse momentum

\end{keyword}

\end{frontmatter}

\section{Introduction}
\label{Introduction}

The associated production of invisible particles and hadronic jets is distinctive of a large number of the topical scenarios currently investigated at hadron colliders. 
The prototypical particle that goes undetected in collider experiments is the neutrino. Neutrinos, which are produced in large part of the electroweak processes, interact only through the weak force, and their passage is inferred indirectly from the imbalance of the total transverse energy of the final state particles. Final states including neutrinos and quarks are of large interest in a variety of measurement or searches; a notable example is the search for the associated production of an Higgs boson and a $Z$ boson, in events where the Higgs boson decays to a $b \bar b$ pair, and the $Z$ boson decays to two neutrinos\,\cite{CDF_nunubb, Abazov:2009jf}. 

Many physics scenarios beyond the Standard Model predict the existence of particles with such a low interaction cross section to result in an imbalance in total transverse energy. Among these, the quest for dark matter is a crucial topic in high energy physics field. Searches at colliders can take advantage of jets from initial state radiation to explore direct dark matter production\,\cite{Fox:2011pm}. Recent experimental work suggest that these kind of searches can provide complementary information with respect to direct experiments on the mass and cross section of the hypothetical dark matter particles \cite{Aaltonen:2012jb, Chatrchyan:2011nd}. 
Another example is Supersymmetry, which provides a whole new spectrum of particles, the lightest of which have to be stable, neutral and non strong-interacting. Events with production of squarks or gluinos are expected to include strong-interacting SM particles in the final state, to conserve the color charge, together with lightest supersymmetric particles, resulting in missing transverse momentum plus jets signatures. 

In all instances, the presence of undetected particles is signaled in collider detectors by an imbalance in the transverse momentum of the final products, which is generally measured with the use of calorimeters and is traditionally referred to, in literature, as ``missing transverse energy" ($\met$). The presence of large missing transverse energy is a distinctive feature of all the signatures we have cited, allowing to significantly reduce the contribution of the generic production of strong-interacting particles through the more mundane quantum chromodynamics, \,({\it i.e.}\,QCD background). Still, the presence of QCD is unavoidable due to calorimeter resolution effects. 

The challenging and hostile environment for invisible particle detection at hadron colliders thus calls for the need of innovative detection procedures.
In this work we analyze the performance of a novel invisible particle detection technique, which is based on track information for evaluation of undetected particles' momentum.

New techniques for this purpose are needed in particular for kinematics in which the invisible particles in the final state carry low transverse momentum, because this imply low $\met$. For example, one of the last supersymmetric scenario not ruled out by LHC data is the one where the next-to-lightest supersymmetric particle is one of the 3rd generation squarks, the stop ($\tilde t$) and sbottom ($\tilde b$), the lightest being the neutralinos\,\cite{Papucci}.
After their pair production, $\tilde t$ ($\tilde b$) would likely decay to a top (bottom) quark and a neutralino $\chi$ that goes undetected. This signature is particularly difficult to investigate if the difference between $M_{\tilde t} (M_{\tilde b})$ and $M_t$ + $M_{\chi}$ ($M_b$ + $M_{\chi}$) is small, because in this case the decay products have low momentum. The fact that experimentalists tend to use very stringent cuts on the event missing transverse energy, in order to reduce the QCD background, drastically reduce sensitivity in this region of the SUSY parameters space (see\,\cite{ATLAS_sbottom} as an example).

Dark Matter searches face the same problem if the mass of the Dark Matter particle is $\ge$100 GeV/$c^2$ because in this case large fraction of the collisional energy becomes mass. Together with the lower production cross section, this makes collider experiments less sensitive than direct detection experiments for Dark Matter masses higher than few tens of GeV/$c^2$, even in the case in which the direct detection bounds are limited by a spin-dependent cross section hypothesis \cite{Aaltonen:2012jb, Chatrchyan:2011nd}. The use of QCD rejection techniques like the one we describe in the following could help extend experimental limits to these unexplored regions, since collider experiments are much less limited by the spin-dependent hypothesis. 

In all these cases, since the expected $\met$ is relatively low, square cuts on $\met$ at trigger level would not succeed to efficiently reject QCD background, maintaining at the same time high efficiency on the signal. Therefore, new techniques to identify, reject and eventually model the QCD background are needed to extend limits in these regions of the phase space.

\section{The CDF II detector}
\label{Detector}

The CDF II detector has been extensively described elsewhere in the literature \cite{CDF}. The detector is cylindrically symmetric around the $p \bar p$ beam line; a magnetic spectrometer consisting of tracking devices inside a 3-m diameter, 5-m long superconducting solenoidal magnet with an axial magnetic field of 1.4 T measures the momenta and trajectories of charged particles. A five layer double-sided silicon microstrip detector (SVX) covers the region between 2.5 to 11 cm from the beam axis. Three separate SVX barrel modules along the beam line cover a length of 96 cm, approximately 90\% of the luminous beam interaction region. Three of the five layers combine an r-$\phi$ measurement on one side and a 90$^\circ$ stereo measurement on the other, and the remaining two layers combine an r-$\phi$ measurement with small angle stereo at $\pm$1.2$^\circ$. The typical silicon hit resolution is 11 $\mu$m. Additional Intermediate Silicon Layers (ISL) at radii between 19 and 30 cm from the beam line in the central region link tracks in the SVX to hits in the Central Outer Tracker (COT), a 3.1-m long
open-cell drift chamber \cite{COT} which occupies the radial
range 40-137 cm. Eight superlayers of drift cells with 12
sense wires each, arranged in an alternating axial and
$\pm 2 ^\circ$ pattern, provide up to 96 measurements of the track
position. The position resolution of a single drift time measurement is about 140 $\mu$m. Full radial coverage extends up to
$|\eta| <$ 1 for the COT and up to $|\eta| <$ 2 for the silicon detectors.

Outside the tracking systems and the solenoid, segmented calorimeters with projective geometry are used to reconstruct electromagnetic (EM) showers and jets. The
calorimeter is segmented radially into lead-scintillator
electromagnetic \cite{emcal} and iron-scintillator hadronic \cite{hadcal}
compartments, both segmented into towers, each covering a small range of pseudorapidity and azimuth, and in full cover the entire 2$\pi$ range in azimuth and the pseudorapidity regions of $|\eta|<$1.1 (central) and 1.1$<|\eta| <$ 3.6 (plug). The transverse energy $E_T = E sin\theta$, where the polar angle $\theta$ is calculated using the measured $z$ position of the event vertex, is measured in each calorimeter tower. Proportional and scintillating strip detectors measure the transverse profile of EM showers at a depth corresponding to the shower maximum. 
Electrons are identified in the central EM calorimeter as isolated, mostly electromagnetic clusters that also match with a track in the pseudorapidity range $|\eta|<$ 1.1. The electron transverse energy is reconstructed from the electromagnetic cluster with precision $\delta$($E_T$ )/$E_T$ = 13.5\%/$\sqrt{E_T(GeV)} \, \oplus$ 2\%, where the $\oplus$ symbol denotes addition in quadrature. Jets are identified as electromagnetic and hadronic calorimeter clusters using the {\sc jetclu} algorithm \cite{jetclu} with a cone size of 0.4, and their energies are corrected for the calorimeter non-linearity, losses in the gaps betwen towers, multiple primary interactions, underlying event  and out-of-cone losses \cite{jetcorr}. 

Drift chambers located outside the central hadronic
calorimeters and behind a 60 cm thick iron shield detect
muons with $|\eta| <$ 0.6, while muons in the region between 0.6 $<|\eta| <$ 1.0 pass through at least four drift layers lying in a conic section outside of the central calorimeter. 
In both cases, muons are identified as isolated tracks in the COT that extrapolate to track segments in one of the four-layer stacks. 
Two Cerenkov Luminosity Counters (CLC) in the forward region measures the average number of interactions per bunch crossing. 

The CDF trigger system is structured in three levels.

Level 1 trigger uses informations from the COT, the calorimeters
and the muon chambers. The {\it eXtremely Fast Tracker} (XFT) processor reconstructs the charged tracks
using the hits from the axial layers of the COT with reduced resolution. Electrons, photons
and jet candidates are identified imposing the presence of energy in the single
towers of the calorimeters above the threshold values. The value of the sum
of the energy released on all the towers is used for the selections based on
the total transverse energy and the missing transverse energy.
The transverse projection of the tower energies (see Sect.\,\ref{met_mpt}) are calculated with the assumption that the event primary vertex is located at $z\, = \,0$. 
The missing energy at L1 has poor resolution, due to a limited available information and the need to make a fast decision, and is usually underestimated.

At Level 2, the XFT tracks with $\geq$ 4 hits in
SVXII and $p_T >$ 2 GeV/c are reconstructed taking advantage of
the additional information from the silicon detector. 

Algorithms can be applied to
trigger electrons, photons or jets: the energy for clusters of adjacent towers,
and the information from the detectors of maximum expansion of the shower
in the electromagnetic calorimeters are now available.

Level 3 uses complete informations supplied from the various detectors. 

In particular, the
tracking is completed executing the three-dimensional reconstruction of the
trajectories in the volume $|\eta| <$ 2, and more detailed algorithms reconstruct
the energy in the calorimeters.
The information provided by the cluster finding algorithm at the trigger level can be considered as a first-order jet reconstruction. At this level, a looser time constraint enables to exploit the full detector segmentation for a better jet energy and direction determination.

\section{Evaluation of the invisible particles' transverse momentum}
\label{met_mpt}

Since the discovery of the $W$ boson, the missing transverse energy has been a crucial tool in
the search for new phenomena at hadron colliders.
Particle colliding at hadron colliders have equal and opposite
momenta; therefore, the total vector momentum sum in an event should be zero. The
hard collision happens between the partons of the proton (antiproton), where they
can carry any fraction of the parent proton or antiproton momentum. 
However, in the plane transverse to the beam, the initial partons' momentum - and thus final ones as well - can be treated as zero to a good approximation. 
Any transverse energy imbalance in the detector may indicate that a particle left the detector without interacting with its
material.

Missing transverse energy, $\met$, is defined as the magnitude of the vector

\begin{equation}
 \vecmet =  - \sum_{i} E^i_T \vec{n}_i
\end{equation}

where $E^i_T$ are the magnitudes of transverse energy measured in each calorimetric tower $i$, and  $\vec{n}_i$ is the unit vector from the interaction vertex to the tower in the transverse $(x, y)$ plane. In order to maintain good resolution on both magnitude and direction of the missing transverse energy, the detector calorimeter must be designed to be almost completely hermetic, within the mechanical allowance. 

The $\met$ resolution is parametrized in terms of the total scalar energy deposited in the calorimeter ($\Sigma E_T$), and is measured with minimum bias events, {\it i.e.} events collected requiring hits in both CLC counters, with no calorimeter requirements. No significant $\met$ is expected in minimum bias events. A fit to these data yelds:

\begin{equation}
 \sigma(\met_x) =  - 0.60 + 0.74\sqrt{\Sigma E_T}
\end{equation}

Similarly to the case of the missing transverse energy, it is possible to define a missing transverse momentum $\vecmpt$ using the charged particle spectrometer, as the negative vector sum of the charged particles momenta $\vec{p}_T$:

\begin{equation}
\vecmpt = - \sum_{tracks} \vec{p}_T
\end{equation}

In events where only charged particles and undetected particles are produced, the $\vecmpt$ is highly correlated in module and direction to the undetected particle(s) momentum, and thus provide  a way to measure their energy with potentially better resolution than $\vecmet$. 
The presence of quarks/gluons in the final state complicates the picture. In the parton shower and hadronization process forming a jet, most particles produced are pions, with a smaller contribution of kaons. 
Due to isospin symmetry, roughly 2/3 of the energy of a jet will be carried by charged particles, which will be measured with both calorimeter and tracking chamber. The $\mpt$ underestimates the undetected particle's energy because it does not take into account the energy carried by the neutral components of the jets, and for the same reason has a worse angular resolution than $\met$. In addition, while calorimeter coverage is usually almost complete, the spectrometer coverage in collider detector is generally far more limited.
For these reasons, $\vecmpt$ cannot substitute the role of the $\vecmet$ as a tool for the measurement of momentum and direction of undetected particles; still, the $\vecmpt$ can provide informations complementary to those given by $\vecmet$. 

In order to reconstruct the composite $\vecmpt$ observable, we use as inputs to its computation the charged particle momentum reconstructed in the spectrometer. The basic track quality criteria for the calculation of $\vecmpt$ are the same used for the reconstruction of the primary vertex at CDF. 
Only tracks with 0.5 GeV/c $< p_T <$ 200 GeV/c, $|\eta| <$ 1.5 and $|Z_{vtx}| <$ 2 cm are used, where $Z_{vtx}$ is the closest approach distance of the track from the primary vertex along the $z$ axis. 
We then classify the tracks in four different categories, on the basis of the number of axial ($N_{COT}^{ax}$, $N_{SVX}^{ax}$) and stereo ($N_{COT}^{st}$, $N_{SVX}^{st}$) COT and SVX layers that have at least 5 hits, together the $\chi^2$ of the track fit. The first category fulfills strong requirements on the track reconstruction in the COT; the second, third and fourth have decreasing requirements on COT compensated by requirements on the $\chi^2$ and on the SVX hit layers. If the first category requirements fail, the second category is checked, then the third and finally the fourth.

\subsection{Data and simulated samples used}

We use for the following studies 5.7 fb$^{-1}$ of data, collected requiring
 the presence of large $\met$ and the presence of two or more energetic hadronic jets.  
At Level 1, the trigger requires at least one trigger tower with $E_T \, \geq \, 10$ GeV, and a $\met \, \geq 28$ GeV. 
At level 2, at least 2 calorimetric clusters with transverse energy above $3\,$GeV and $|\eta| \, \leq \, 3.6$ are required. The $\met$ is recomputed with the additional available informations; again, $\met$ must be greater than 28 GeV.
Finally, the Level 3 performs the complete $\met$ determination, requiring it to be greater than 30 GeV. 

Events containing large $\met$ can originate from non-collisional sources, such as cosmic or beam-halo muons
passing the detector or noisy/dead calorimeter cells causing energy imbalance. These
types of events are removed by requiring that the event observables indicate a presence
of inelastic collision with large energy transfer, such as the presence of at least one high
quality primary vertex in the collision.
Additional requirements are also imposed to remove events consistent with beam-halo
muons traversing the detector or those caused by noisy calorimeter cells. We require the corrected $\met$ to be larger than 50 GeV, and we veto events containing at least one isolated electron or muon. We require the transverse momentum of the jet $p_T(i)$ (where $i$ runs over the total number of jets) to be greater than 40 GeV for dijet events, and greater than 30, 20 and 15 GeV respectively for ($i \, =$ 1, 2), ($i \, =$ 3, 4, 5) and ($i \, > $ 5) for all other jet multiplicities. All jets must satisfy $|\eta| < $ 2.4.  With these selections, the QCD multijet production accounts for more than 95\% of the data \cite{Aaltonen:2011na}; for this reason, we will mostly use the data themselves to show the feature of the QCD background. 

On the other hand, in order to describe the qualitative features of physics processes giving rise to undetectable particles in the final state, we use detailed Monte Carlo (MC) simulation interfaced to the detector response simulation. For simplicity, we restrict ourselves to the modeling of the $W/Z$ boson plus jets production, and top quark pair production. As an example of interesting signal, we use here the associated production of a Higgs ($H$) boson together with a $W$ or a $Z$ boson production $WH/ZH$, where the Higgs boson decays to $b \bar b$ and the $Z$ ($W$) boson decays to $\nu \bar \nu$ ($\ell \nu$).
The associated production of a $W$ or $Z$ boson and jets is simulated using the {\sc alpgen} \cite{alpgen} program, interfaced with parton-shower model from {\sc pythia}\, \cite{pythia}. A matching scheme is applied to avoid double counting of partonic event configurations \cite{matching_scheme}. The samples are normalized to the inclusive cross sections \cite{WZ_crosssections}, scaled by 1.3 to account for next-to-leading-order corrections. We model $t \bar t$ using {\sc pythia} with  top quark mass equal to 172.5 GeV/c$^2$, which is consistent with the current world best estimate of this parameter \cite{Lancaster:2011wr, Galtieri:2011yd} and normalizing its contribution to the $t \bar t$ NLO cross section \cite{ttbar_cross_section}). Associated production of a Higgs ($H$) boson together with a $W$ or a $Z$ boson production $WH/ZH$ is modeled with the {\sc pythia} MC generator, assuming a Higgs boson mass equal to 115\,GeV/c$^2$. In order to test MC prediction for QCD events, we generate dijet and trijet events with {\sc pythia}. 

For all the simulated samples above, the detector response is modeled by a {\sc geant}\,\cite{GEANT}-based simulation of the CDF detector.

\subsection{\met and \mpt distributions in QCD multijet events}

\begin{figure}
    \includegraphics[width=7cm]{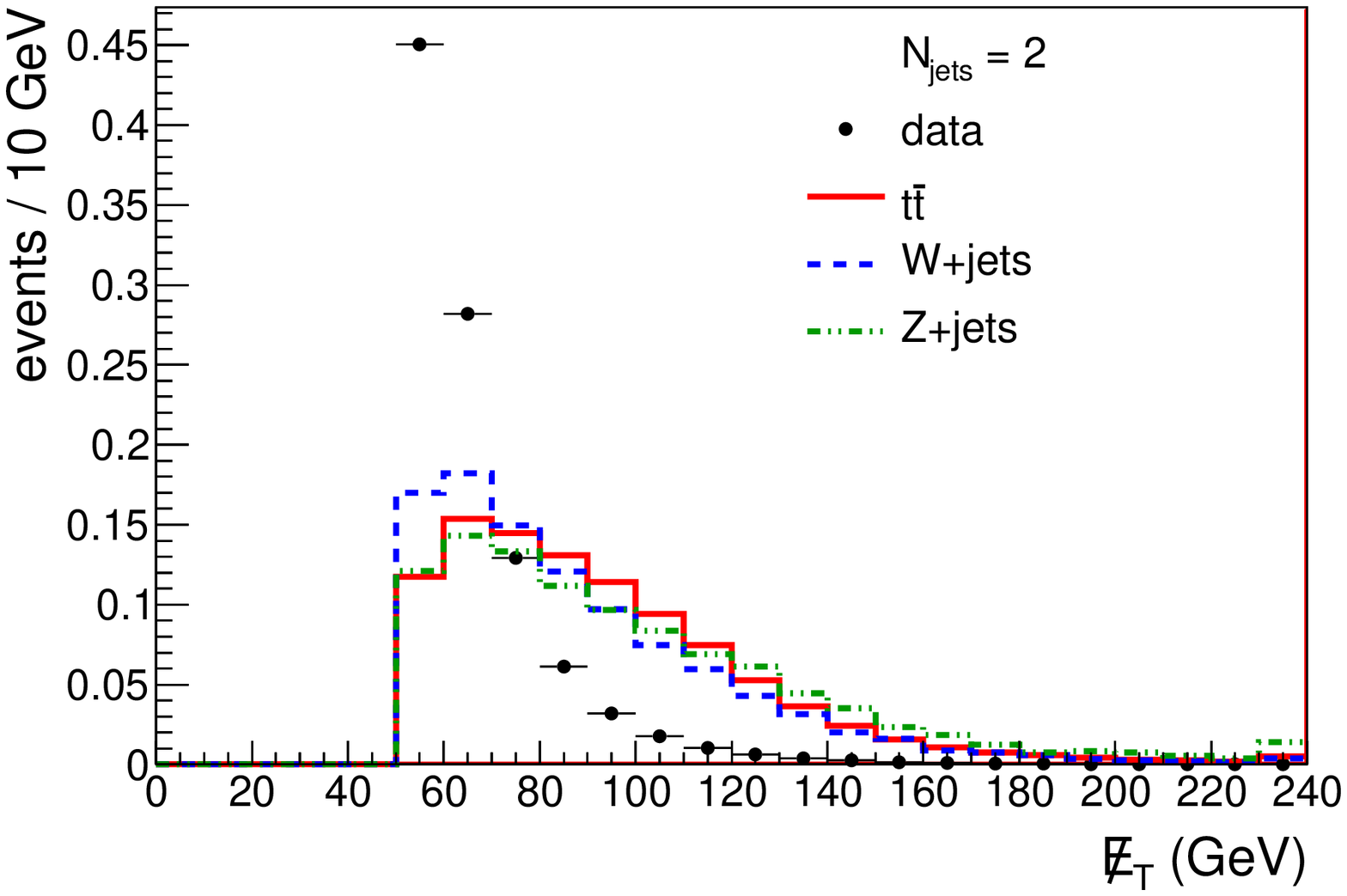}
    \includegraphics[width=7cm]{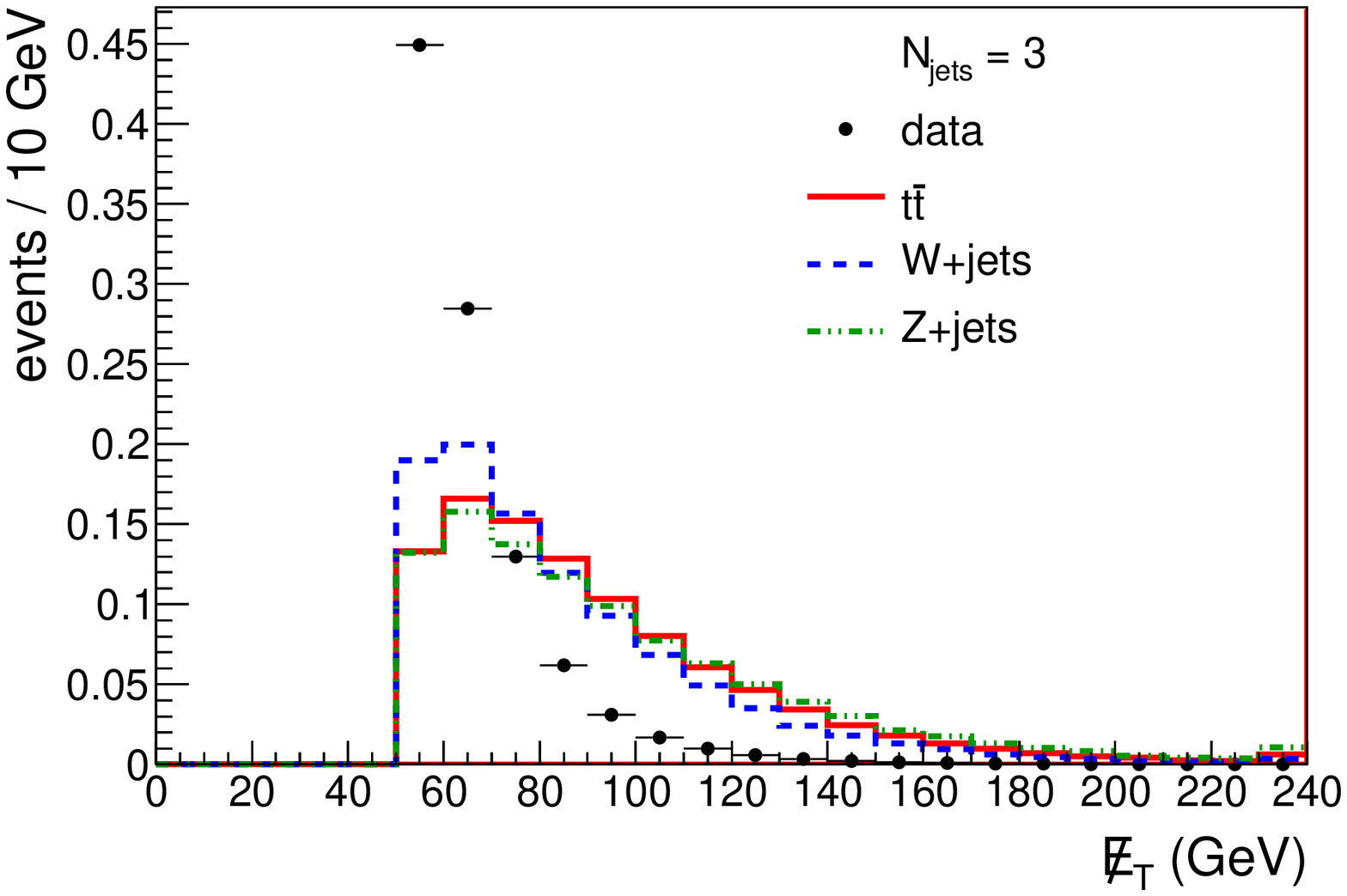}		
    \includegraphics[width=7cm]{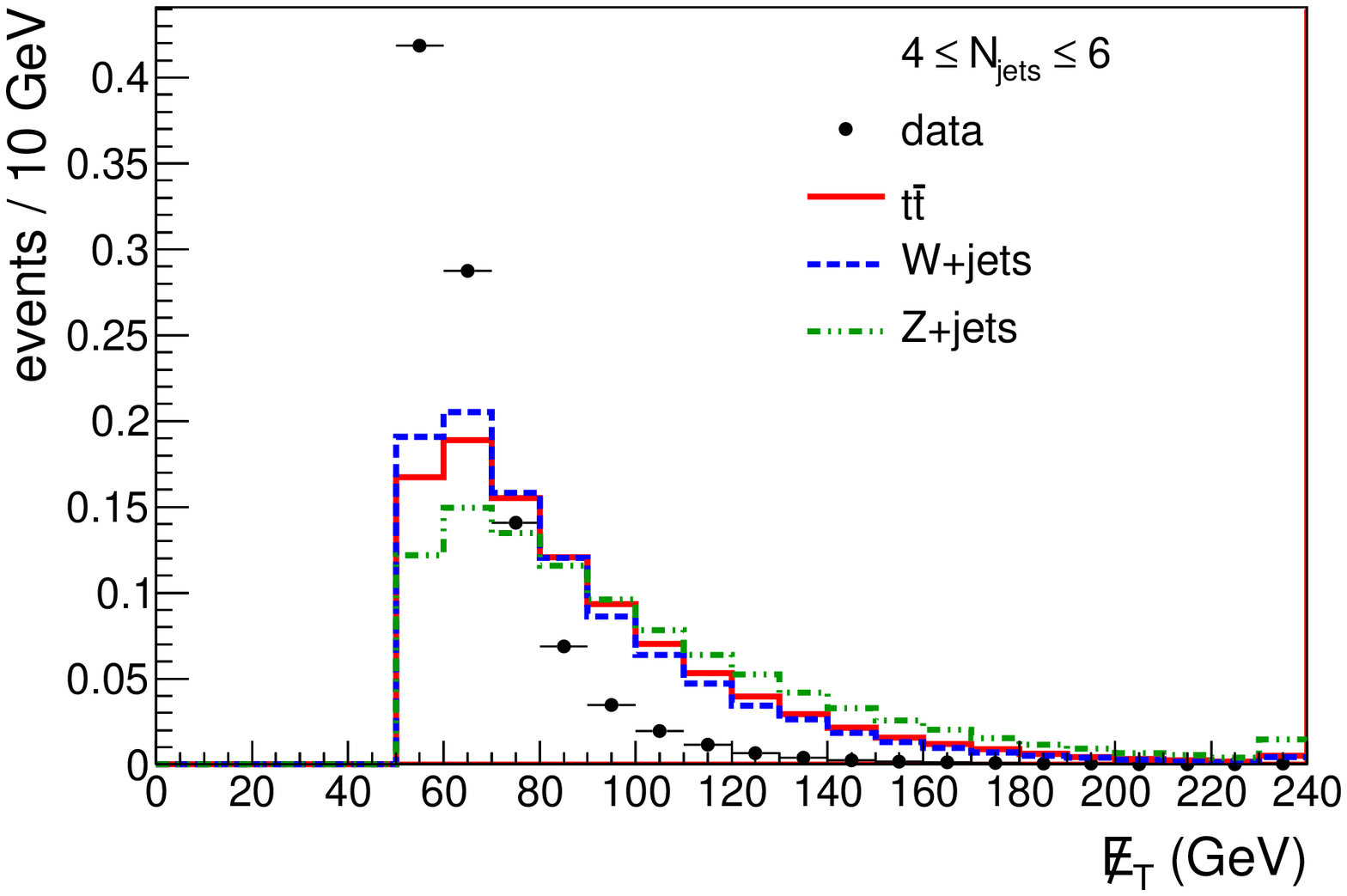}
    \includegraphics[width=7cm]{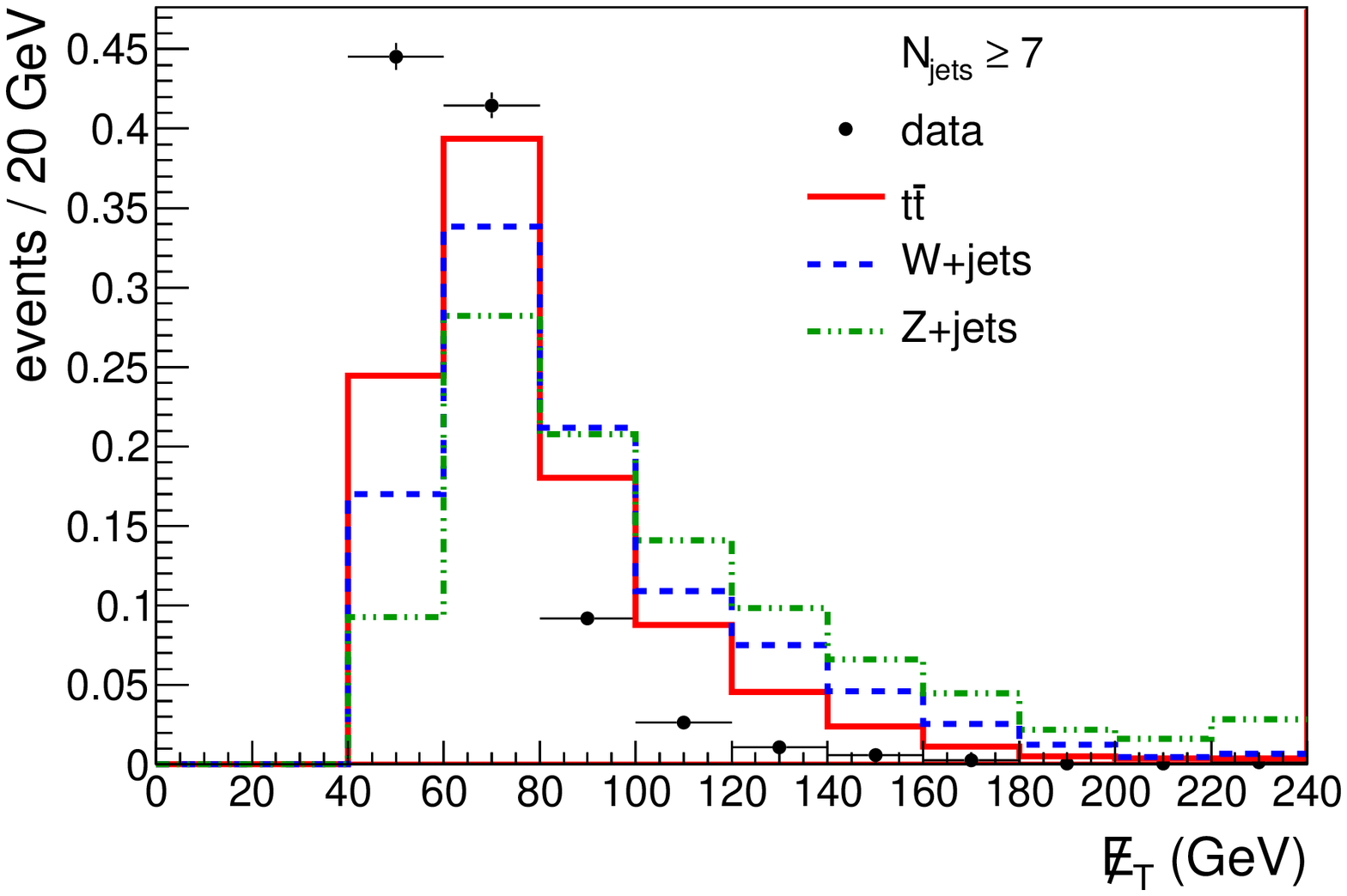}
    \caption{\label{fig:MET} The distribution of $\met$ for data and major Standard Model processes giving rise to neutrinos. The events are selected with $\met > 50 GeV$ and number of jet $N_{jet}$ varying from 2 (top-left plot) to 7 or more (bottom-right plot). All distributions are normalized to unit area.}
\end{figure}

The generic production of strong-interacting particles by quantum chromodynamics interactions between quarks and gluons is the most probable source of hadronic jets in the particle colliders. Because of this, it is also the greatest source of backgrounds for searches which do not require leptons in the final state. 
QCD multijet production does not include undetectable particles, apart from small contributions from heavy quarks semileptonic decays, which produce neutrinos. Missing transverse energy in these events, if correctly measured, is thus expected to be zero. This allows to significantly reduce the QCD multijet background for signatures including undetectable particles. Nevertheless, calorimeter resolution as well as non linearity and non hermeticity have to be taken into account, resulting in a smearing of the $\met$ distribution. As a result of this, the largest part of the events which pass a $\met >$ 50 GeV requirement, which is loose enough to be used with low final momentum signatures (see Sect. \ref{Introduction}), is mainly constituted by QCD multijet.
Furthermore, this contribution is difficult to model, especially in topologies consisting of large jet multiplicities. For this reason, analysis looking for $\vecmet$+jets generally take advantage of data-driven techniques to model QCD contribution.  
In Fig. \ref{fig:MET} are shown the $\met$ distributions for events with $\met >$ 50 GeV, for jet multiplicities ranging from 2 to $\geq$\,7; as can be seen, the QCD multijet is concentrated to lower $\met$values, while MC simulations of $t \bar t$, $W$+jets and $Z$+jets have, except in full hadronic decay modes, real $\met$ due to neutrinos.

Again, the data themselves are taken as a representation of the QCD background. The distribution shows that physics processes giving rise to high-$p_T$ neutrinos display a $\met$ distribution with a pronounced tail even after a given cut on the magnitude of the missing transverse energy - as expected.

\begin{figure}
    \includegraphics[width=7cm]{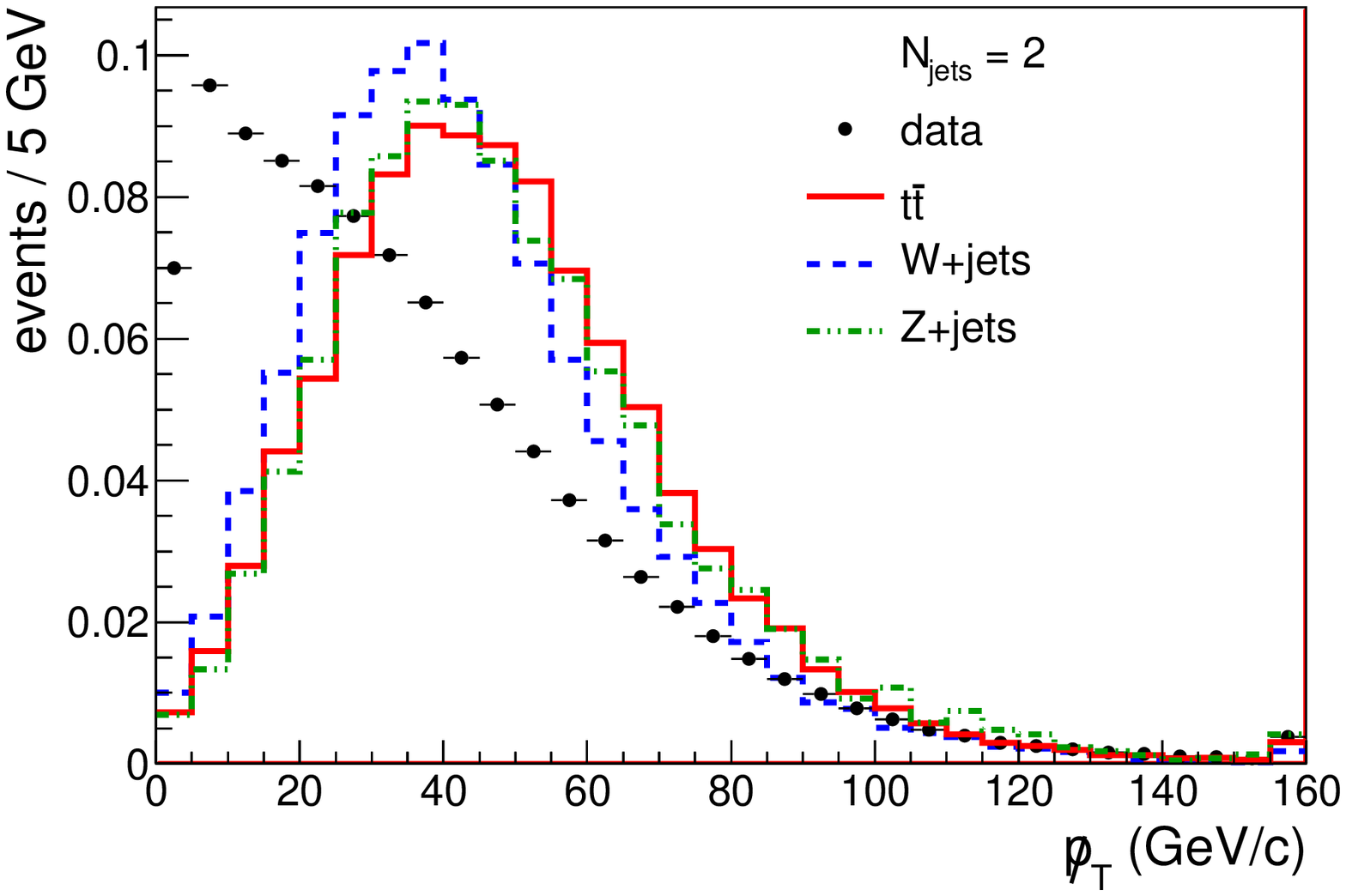}
    \includegraphics[width=7cm]{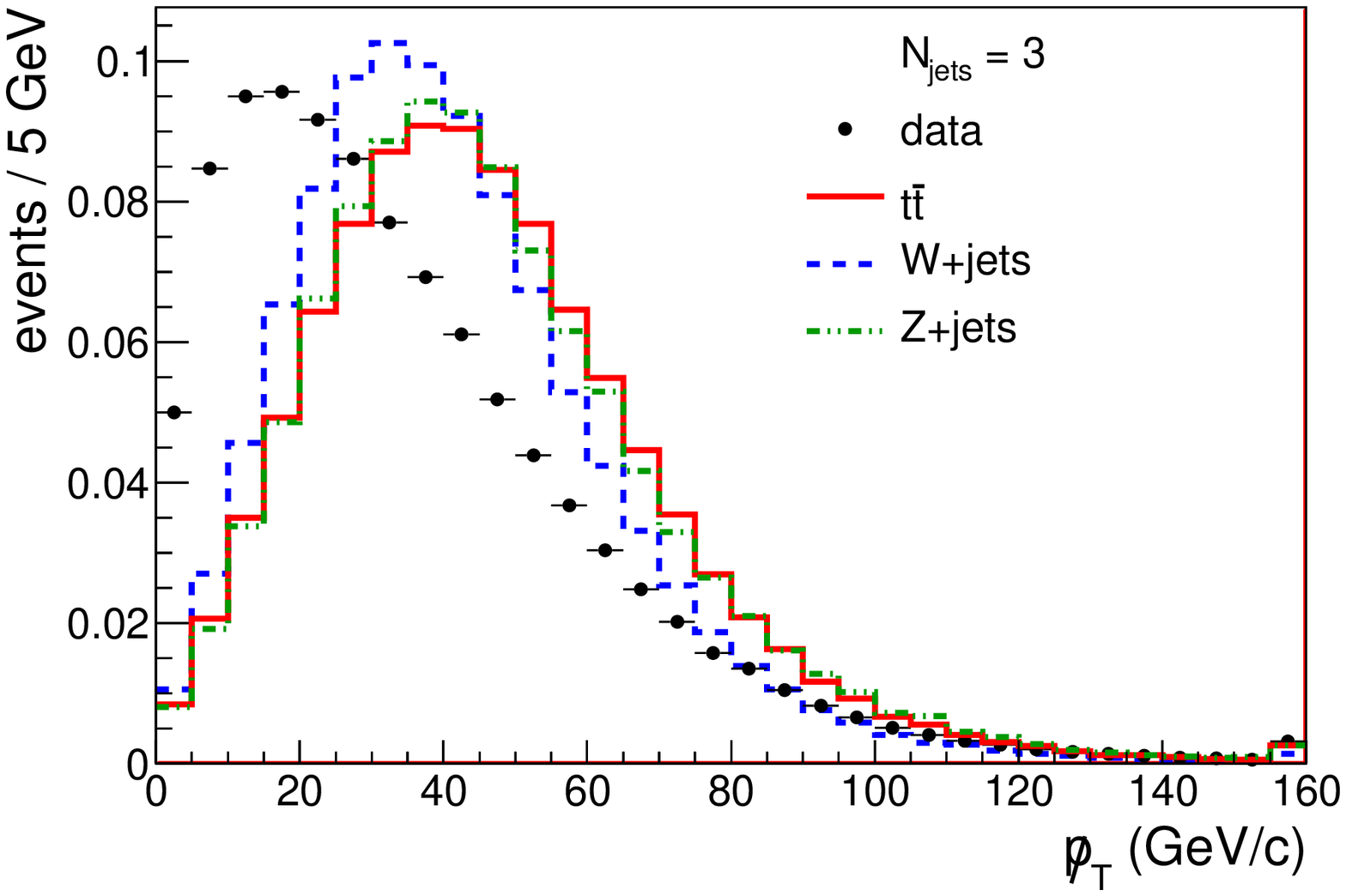}		
    \includegraphics[width=7cm]{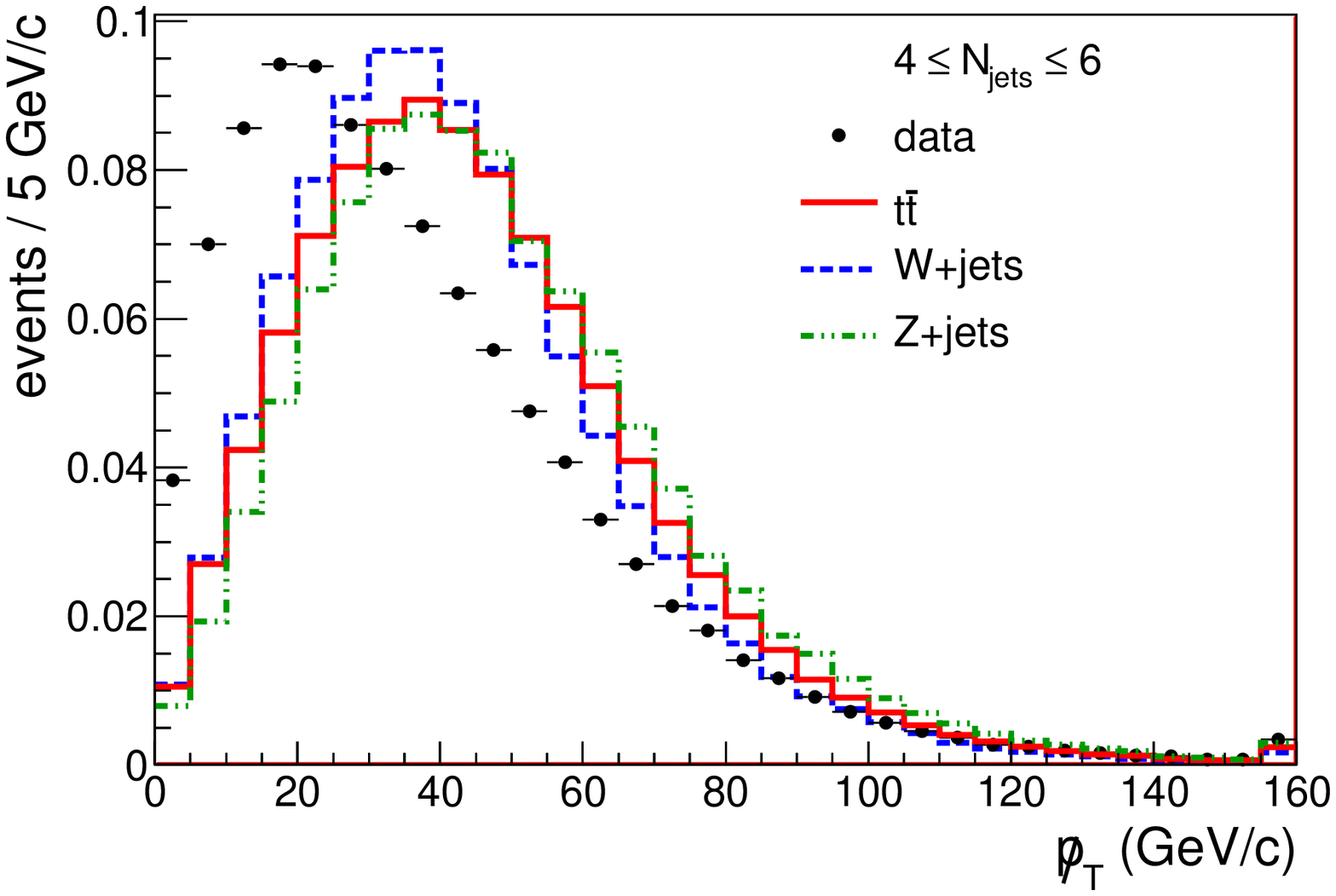}
    \includegraphics[width=7cm]{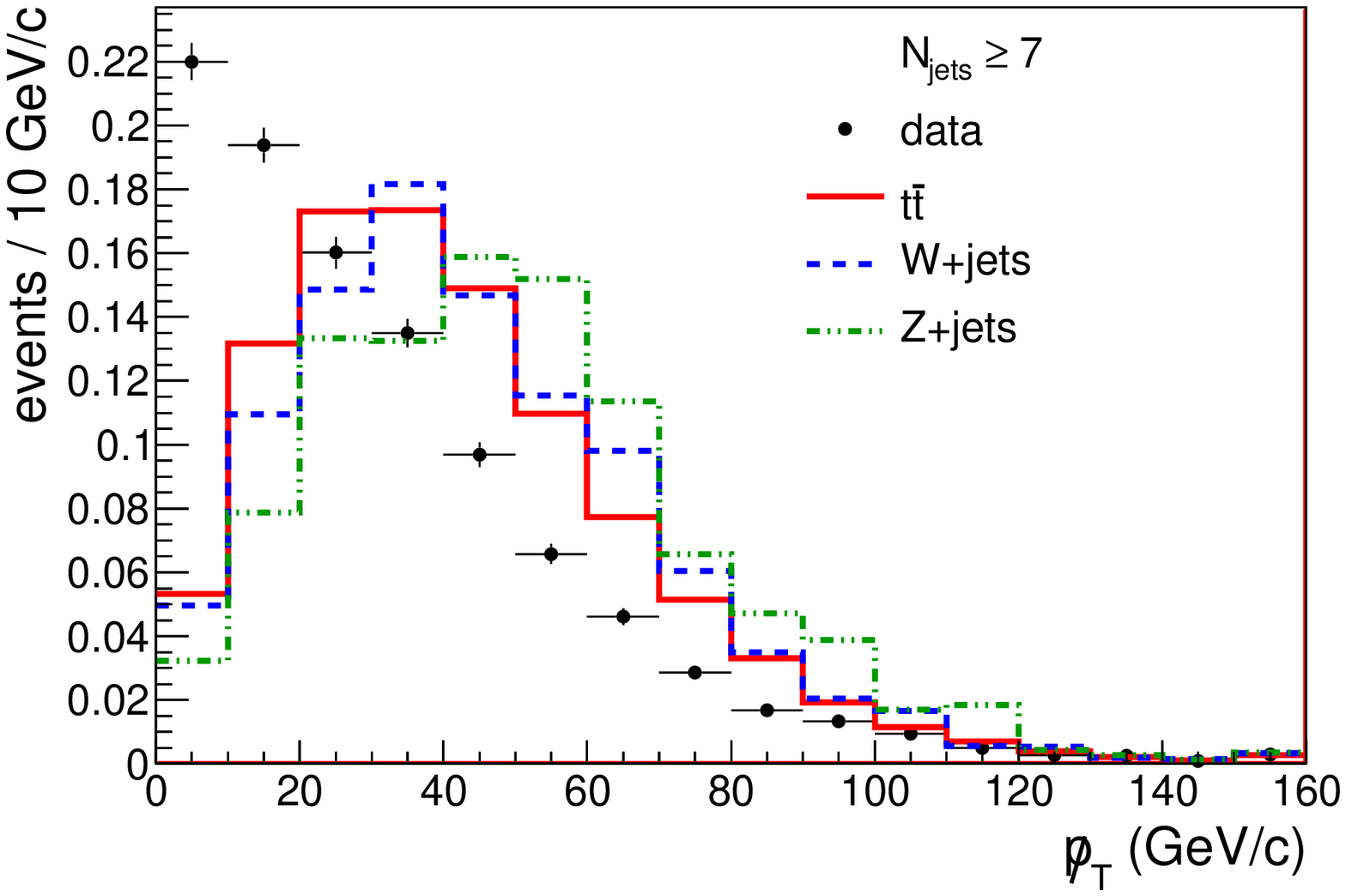}
    \caption{\label{fig:MPT}    The distribution of $\mpt$ for data and major Standard Model processes giving rise to neutrinos. The events are selected with $\met > 50 GeV$ and number of jet $N_{jet}$ varying from 2 (top-left plot) to 7 or more (bottom-right plot). All distributions are normalized to unit area.}
\end{figure}

Figure\,\ref{fig:MPT} shows the $\mpt$ distribution for the same physics processes, after the same event selections. It can be noted that the $\mpt$ distribution for events containing real neutrinos in the final state tend to peak close to the $\met$ values for the same processes. On the other hand, the data - representing QCD multijet production - peaks at much lower values, in a manner compatible with the interpretation that $\mpt$ in QCD events arises from random fluctuations of the charged particle component of each jet in the event, and that they tend to cancel out in the simple dijet topology.
The bottom line is that the $\mpt$ clearly shows a notable separation power between events where $\met$ arises because of resolution effects, and physics processes where $\met$ is expected to signal the passage of undetectable particles.

\section{Angular correlations between \vecmet and \vecmpt}
\label{Angular correlations}

 For events with real undetected particles, $\vecmet$ is a good measurement of the vectorial sum of those particles four-momenta , and so is the $\vecmpt$ which thus will tend to be parallel to the $\vecmet$. However, in QCD events the two variables have very different origin, which is reflected in their different expected angular correlation. 
 We will describe this aspect in details, starting from the simplest topology.

\subsection{Dijet events}

\begin{figure}
  \begin{center}
    \includegraphics[width=6cm]{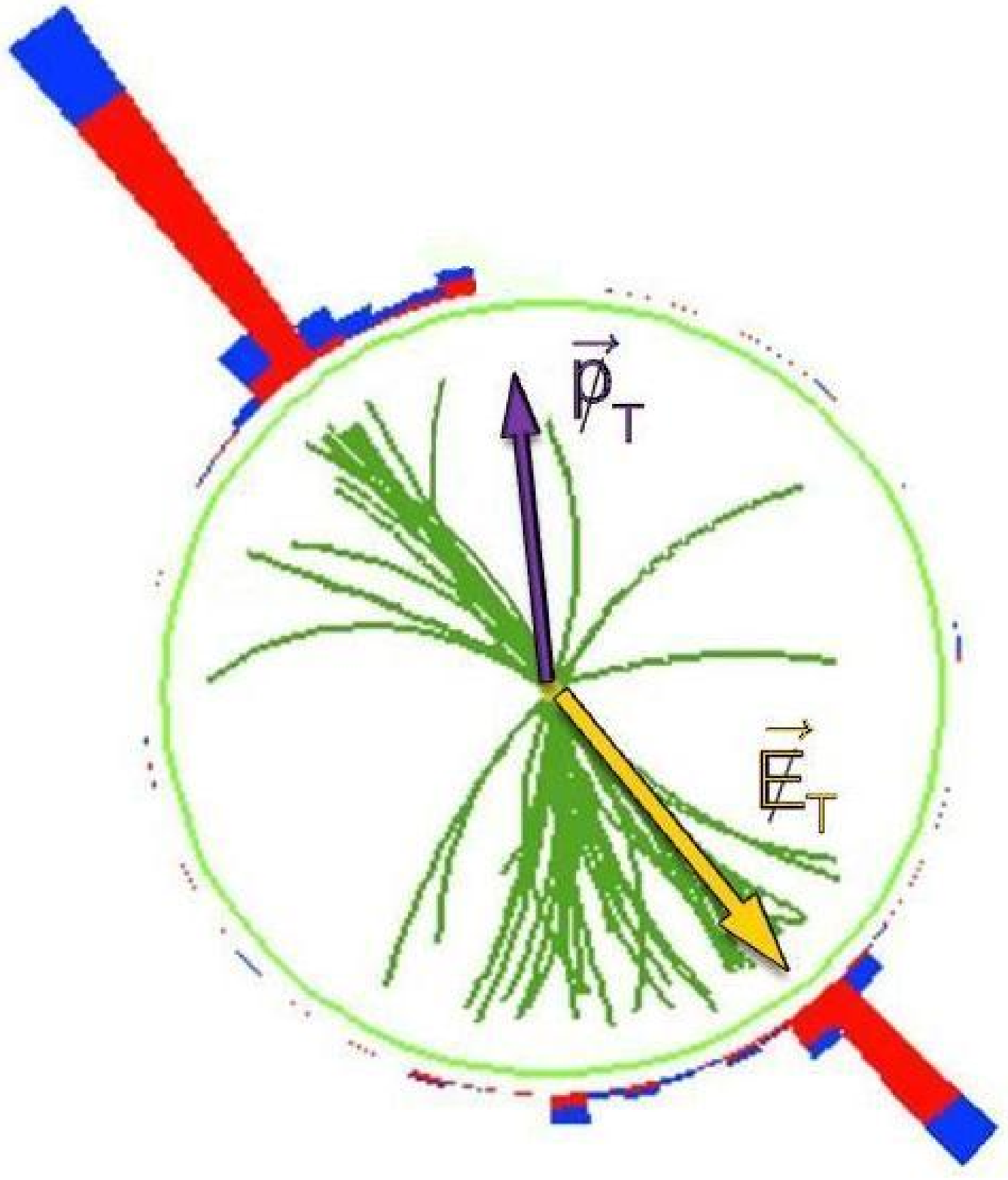}
    \includegraphics[width=6cm]{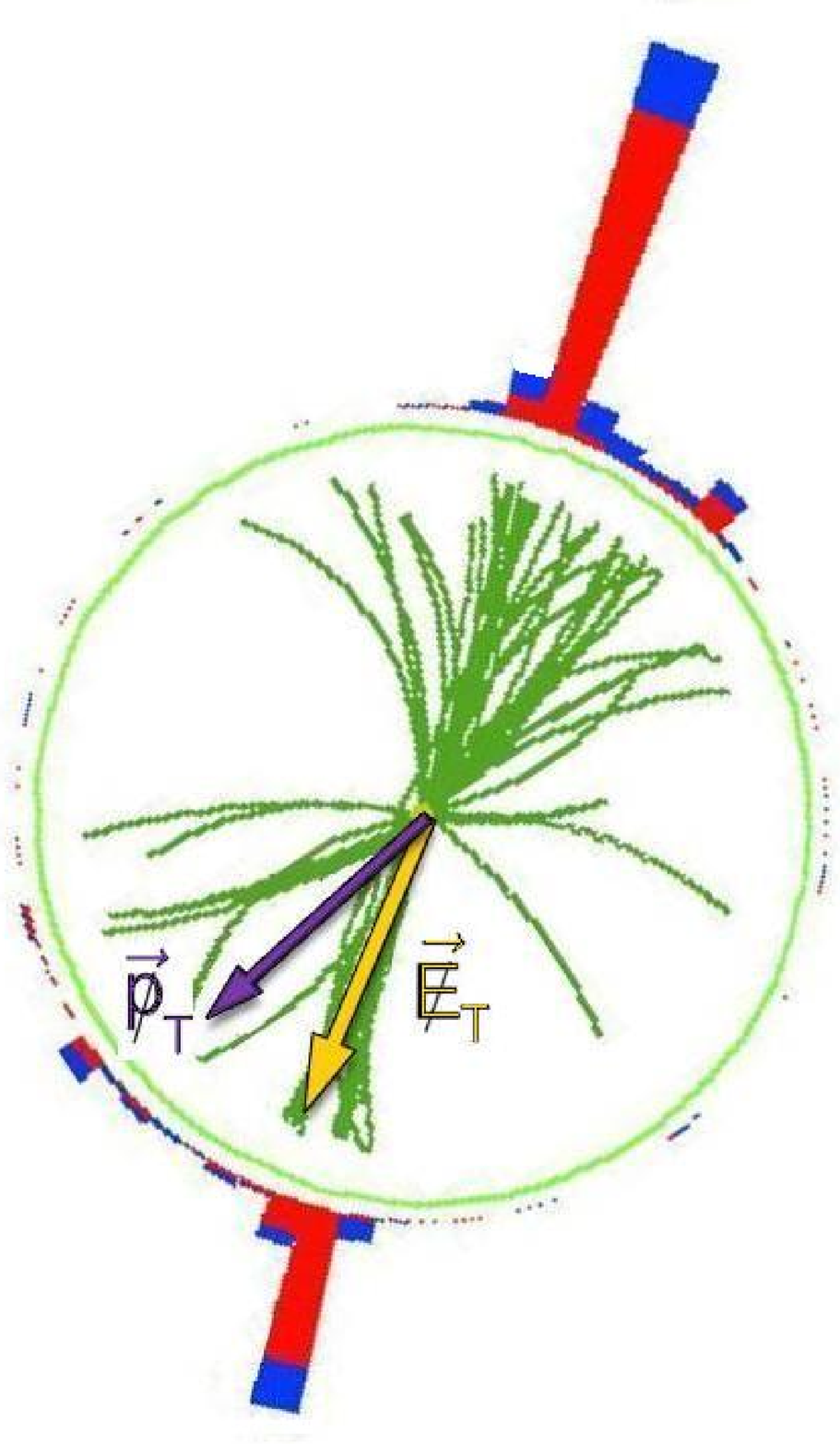}	
   \caption{\label{fig:sketch} Schematical representation of the energy and momentum flow in a QCD dijet event. In both plots, $\vecmet$ is due to calorimetric mismeasurement. The $\vecmpt$, which is due to statistical fluctuation in the vectorial sum of charged particles' momenta in each jet, can arise with equal probability in the correctly-measured jet (top) or in the under-measured jet (bottom). }
  \end{center}
\end{figure}

For QCD events, the energy conservation requires that the vector sum of the jet transverse energies amounts to zero. In dijet events, the 2 jets will have the same magnitude of transverse energy, and will come out back-to-back in azimuthal space. The mismeasurement of jet energies makes the $\vecmet$ align to the jet with less measured energy. The $\vecmpt$ could be present in some amount too, but for different reasons. In the jet fragmentation and hadronization processes a certain number of charged particles are produced inside each jet, and they will be detected by the tracker. The intrinsic fluctuations inherent to the parton shower process will result in large fluctuations in the fraction of energy carried by charged particles inside a jet. The fraction of charged particles for a jet is completely independent from the measured jet energy in the calorimeter. Since is more likely for a jet's energy to be undermeasured than to be overmeasured, the $\vecmet$ will always be aligned to the under-measured jet, while the $\vecmpt$ 
will mainly point to the direction of jet with less energy being carried by reconstructed charged particles. As a net result, in QCD dijet events the $\vecmet$ and $\vecmpt$ directions will be mainly correlated or anticorrelated.  
A schematic representation of the argument above is given in Fig. \ref{fig:sketch}.

\begin{figure}
    \includegraphics[width=7cm]{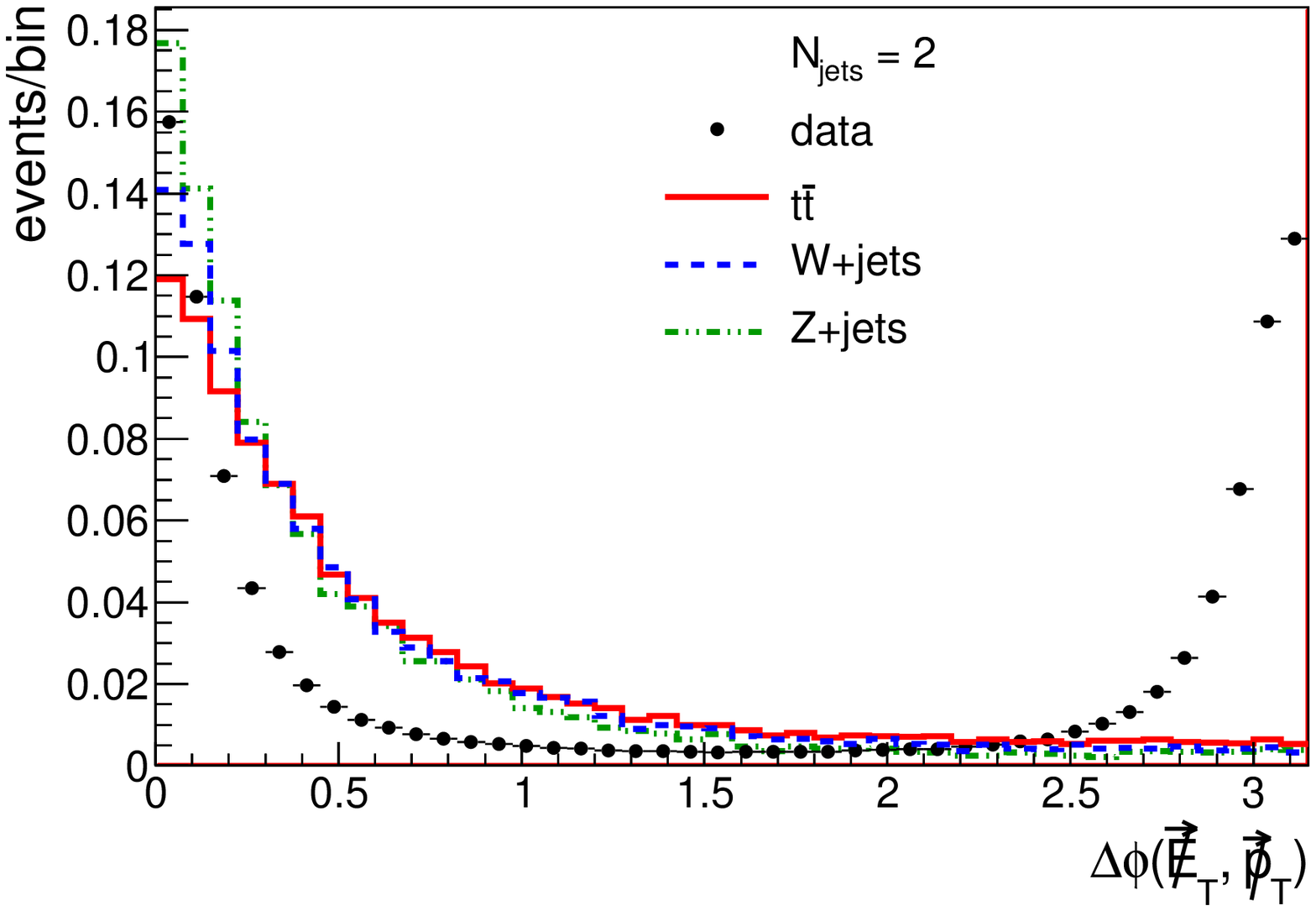}
    \includegraphics[width=7cm]{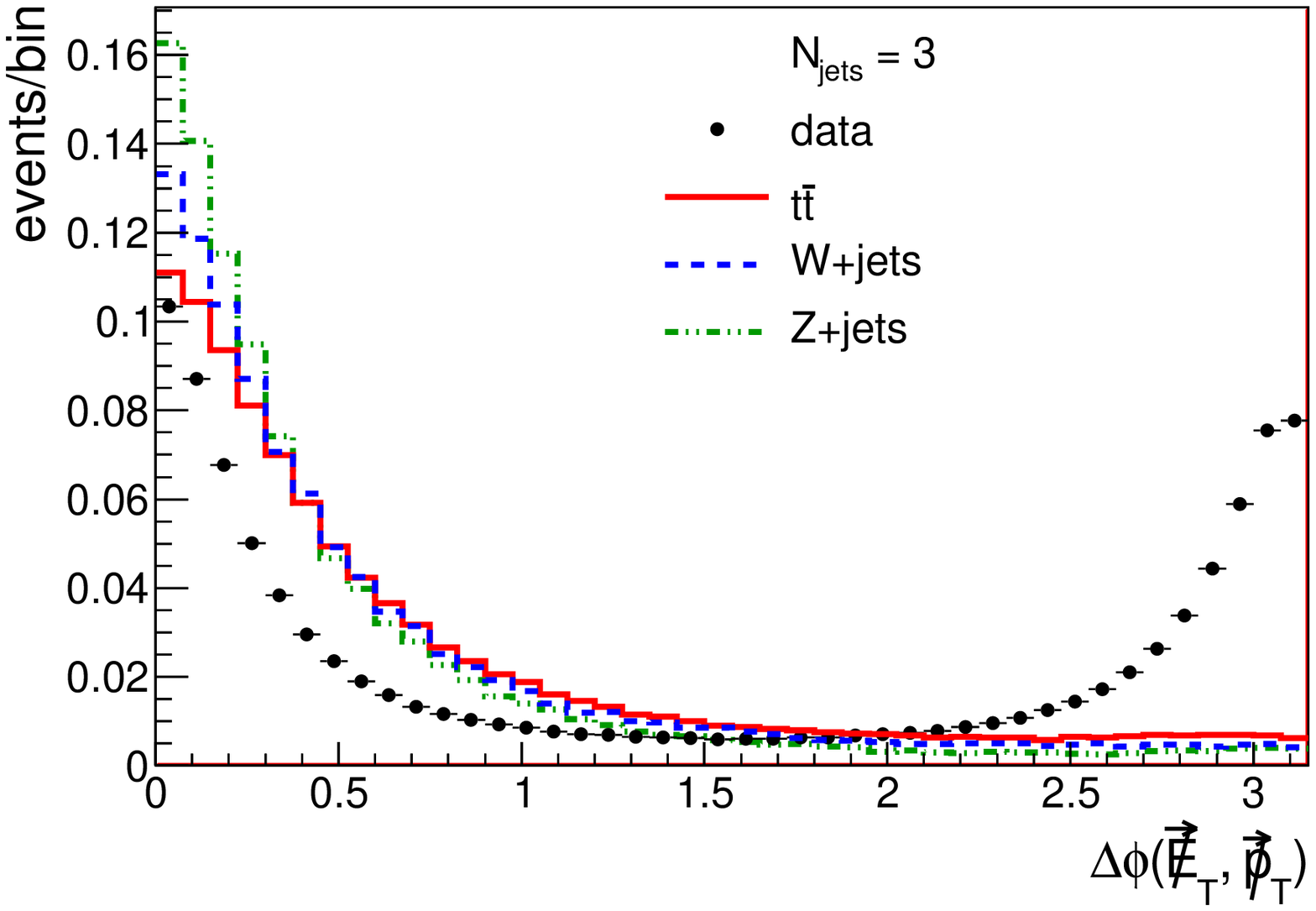}
    \includegraphics[width=7cm]{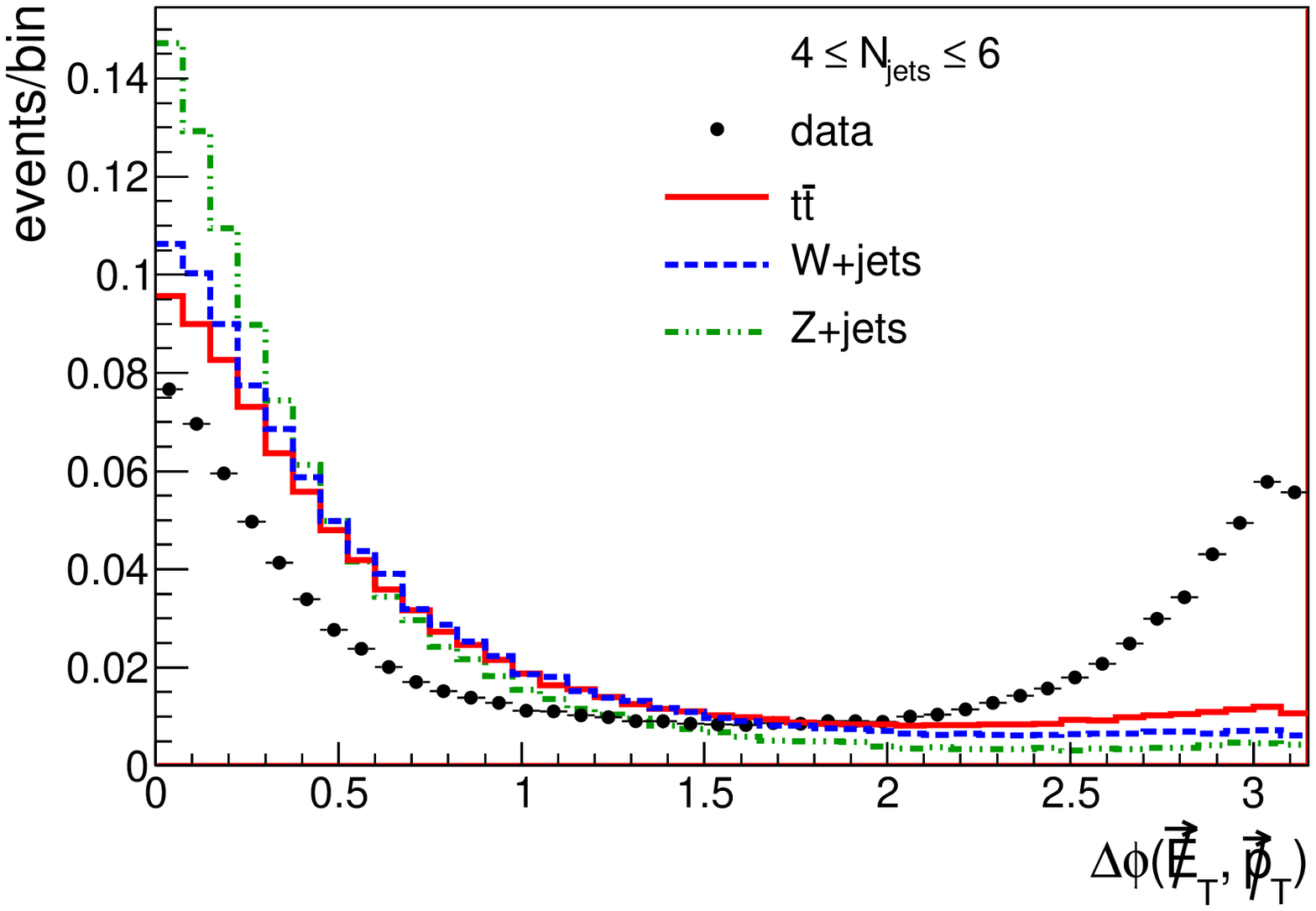}
    \includegraphics[width=7cm]{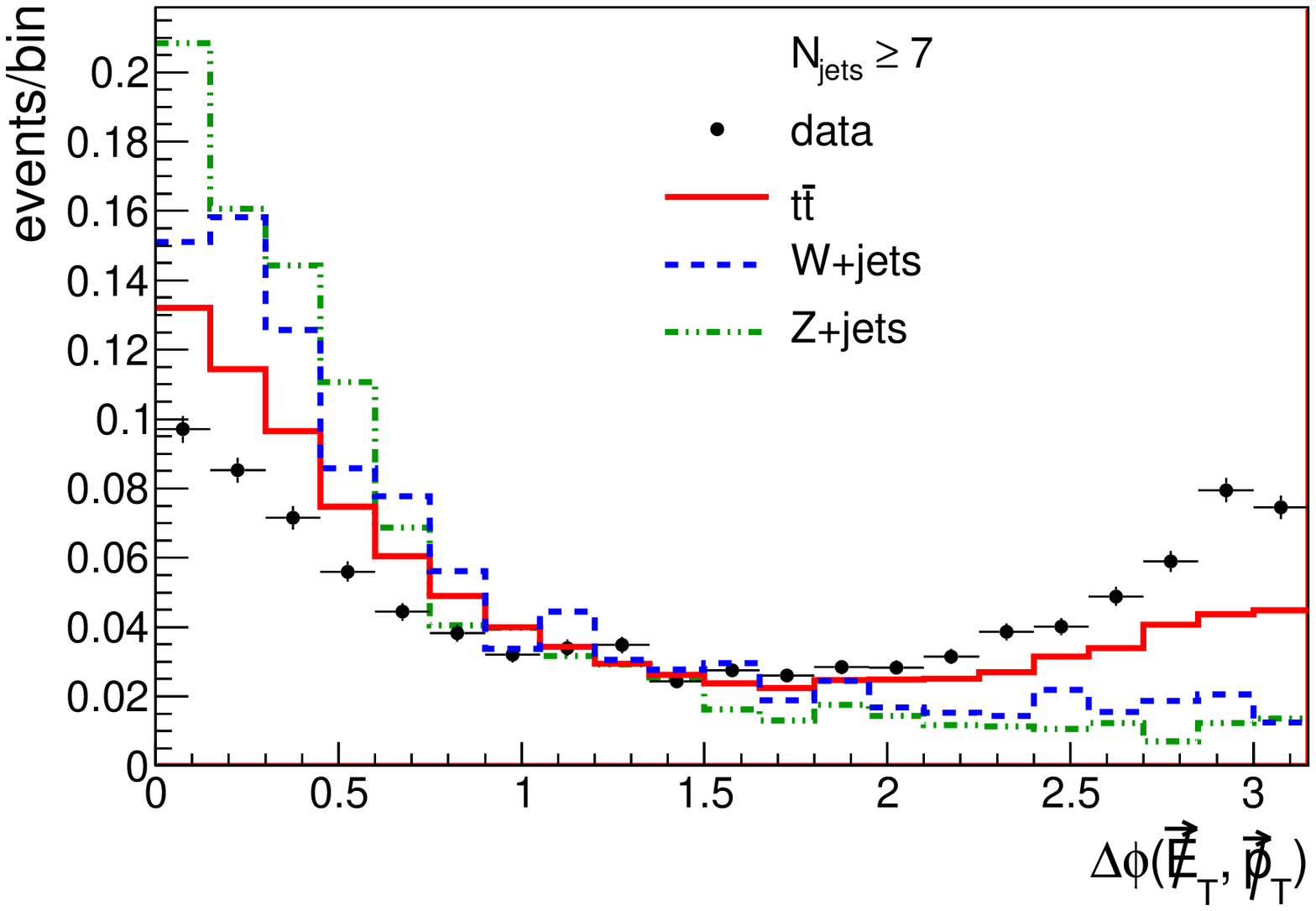}
    \caption{\label{fig:DPhi_METMPT} The distribution of $\Delta \phi(\vecmet, \vecmpt)$ data and major Standard Model processes giving rise to neutrinos. The events are selected with $\met > 50 GeV$ and number of jet $N_{jet}$ varying from 2 (top-left plot) to 7 or more (bottom-right plot). All distributions are normalized to unit area.}
\end{figure}

\begin{figure}
    \includegraphics[width=7cm]{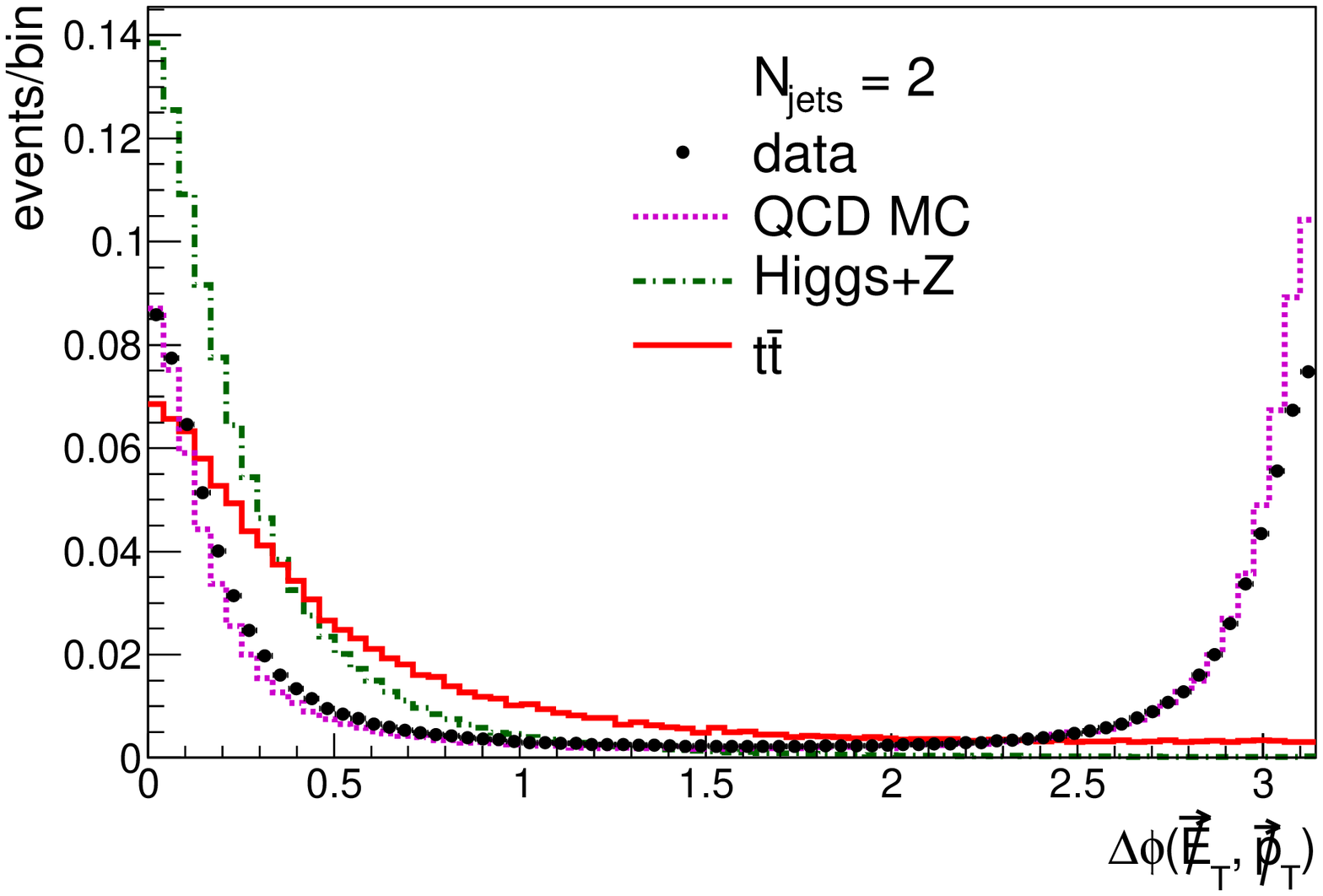}
    \includegraphics[width=7cm]{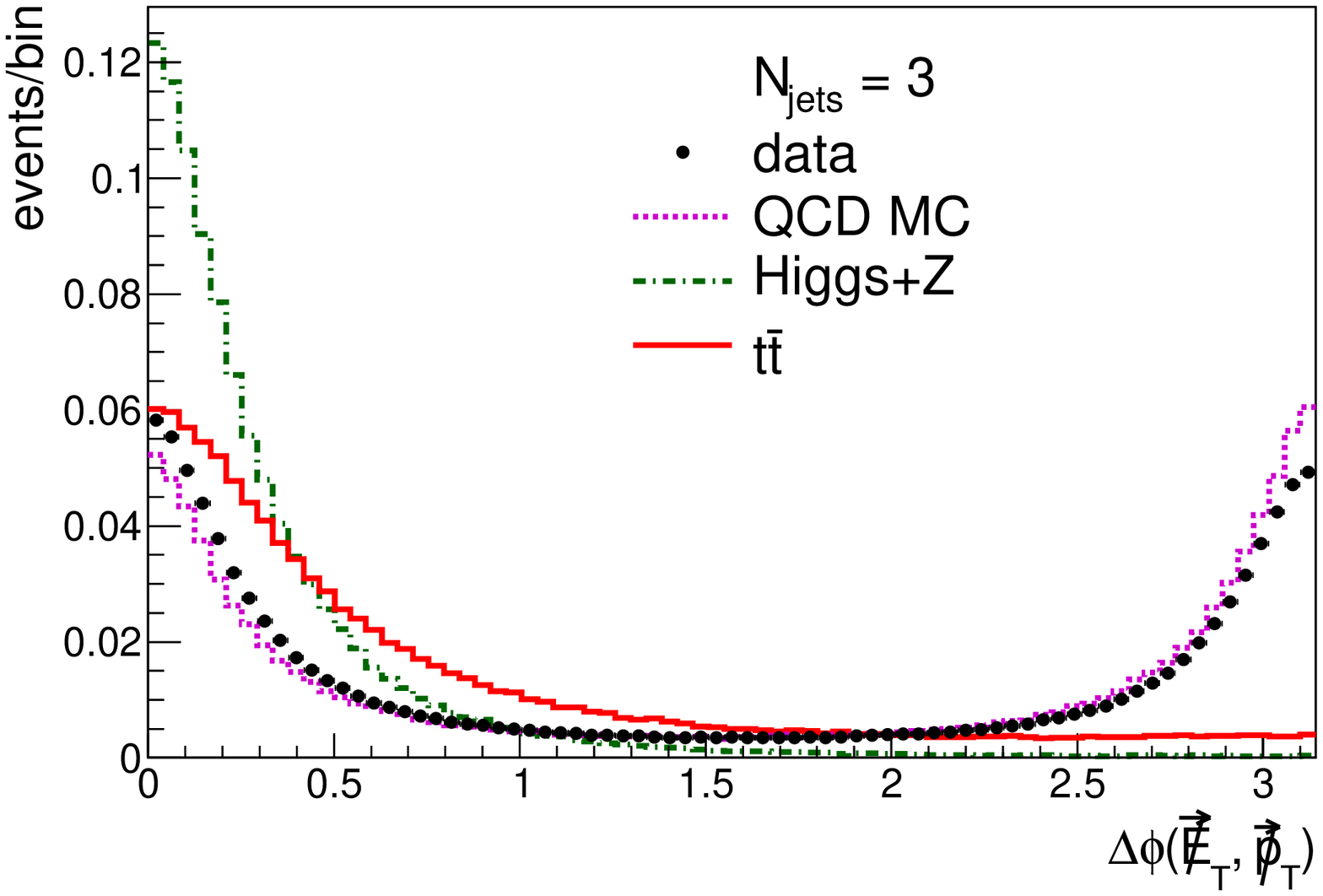}
    \caption{\label{fig:DPhi_METMPT_MC} The $\Delta \phi(\vecmet, \vecmpt)$ distribution in events with 2 (left) and 3 (right) jets. Lines represents QCD, Higgs+W/Z and $t \bar t$ MC, while dots represents data. All plots are normalized to unit area.}
\end{figure}

This particular topology is reflected in the distribution of the azimuthal distance between the two vectors, $\Delta \phi(\vecmet, \vecmpt)$. As shown in  Fig. \ref{fig:DPhi_METMPT}, events containing undetected particles like neutrinos are concentrated near 0, while QCD dijet events have almost equal chance to populate the region around 0 or around $\pi$. This special distribution allows to effectively suppress the QCD contribution in samples with $\met$ and jets. 
Dijet (and 3 jets) events are important because have the most simple kinematic, so that they can be also studied through QCD MC. Fig. \ref{fig:DPhi_METMPT_MC} shows the $\Delta \phi(\vecmet, \vecmpt)$ distributions for data and simulated samples of QCD and events with neutrinos. 
The presence of $b$ quarks in the final state is expected not to alter significantly the situation, because their semileptonic decays would give rise to low-energetic neutrinos.

The effects of fluctuations in the charged fractions of the jet energies can be quantified defining the charged fraction as follows: 

\begin{equation}
p_T^{chgd} = \frac{|\sum_{{Jet_{track}}_i} \vec{{p}_T}_i|}{p_T^{jet}}
\end{equation}

\begin{figure*}
  \begin{center}
    \includegraphics[width=8cm, angle=90]{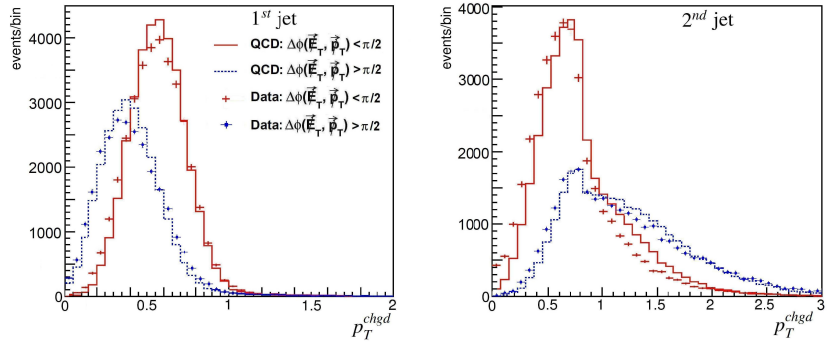}
    \caption{\label{fig:chargedFraction}The $p_T^{chgd}$ distributions for the 1$^{st}$ (left) and 2$^{nd}$ (right) jet in QCD MC, while $\Delta \phi(\vecmet, \vecmpt)$ is in (0, $\pi$/2) or ($\pi$/2, $\pi$); the dots shows the same distributions for data. MC expectations are normalized to data.}
  \end{center}
\end{figure*}

In QCD dijet events the requirement of large $\met$ means that the energy of one of the two jets energy will be largely mismeasured. After ordering the jets according to their $E_T$ starting from the one with the largest energy, the $\vecmet$ will be aligned to the second jet. 
In events where $\vecmet$ and $\vecmpt$ are correlated we expect the $p_T^{chgd}$ of the 2$^{nd}$ jet to be sistematically smaller than in events where they are anti-correlated. The opposite effect will be present for the 1$^{st}$ jet.
This can be observed by looking at the distribution of $p_T^{chgd}$ of the 1$^{st}$ and the 2$^{nd}$ jet, separately for events where $\vecmet$ and $\vecmpt$ are correlated or anti-correlated. As can be seen in Fig.\,\ref{fig:chargedFraction}, the expected results are confirmed, which supports the overall interpretation of the mechanism behind the formation of the $\Delta \phi(\vecmet, \vecmpt)$ distribution for QCD events.

\subsection{Large jet multiplicities}

In energetic QCD multijet events, large part of the energy is carried by the two most energetic jets. These two jets will tend to be roughly back-to-back, and both $\vecmet$  and $\vecmpt$ will have high probability to lie on their axis, while additional jets will arise because of radiation from the initial/final state particles. For these reasons, we expect the exposed $\Delta \phi(\vecmet, \vecmpt)$ distribution features to be recognizable also for QCD multijet events. This is confirmed from the analysis of both data and MC with exactly 3 jets, as is shown in Fig. \ref{fig:DPhi_METMPT} and \ref{fig:DPhi_METMPT_MC}.
The logical step can be extended to higher multiplicities. In fact, for a typical QCD multijet event with 4 or more jets, more than 60\% of the total energy is carried by the two most energetic jets. The distributions of data shown in Fig. \ref{fig:DPhi_METMPT} confirm the validity of this approach up to events with 7 jets or more, that is where CDF statistics run out. The use of MC to cross-check, as for 2 and 3 jets events, is not feasible for higher multiplicities; in facts, because of the high production rate for QCD at a hadron collider and the large statistics needed in order to describe these processes adequately in an analysis of several inverse femtobarns of data, the simulation of an acceptable amount of QCD events is prohibitive.

\section{Study of $\met$ and $\mpt$ as a function of the collider instantaneous luminosity}

The varying instantaneous luminosity at hadron colliders poses a unique challenge to the online selection (triggering) of events characterized by large missing transverse energy and jets in the final state. In fact, larger instantaneous luminosity means more collisions per bunch-crossing. The most common collisions will give rise to quarks or gluons in the final state, and thus jets in the detector. These jets can add up to the missing transverse energy from the hard scatter, or can provide additional missing transverse energy if the energy of any or more than one of those is mismeasured in the detector.

In both cases, the pile-up of multiple scatters will constitute an additional background to the missing transverse energy plus jets signature. Typical solutions proposed to cope with the unsustainable large trigger rates at increasing luminosity, have been to tighten the requirement of the absolute value of the missing transverse energy, or to tighten the requirement in the jet transverse energy, or both. While allowing to control the trigger rates, this procedure often reduces drastically the yields of the interesting signal.

We argue in this section that the spectrometer can be helpful in suppressing the multiple scatter pile-up background in a similar fashion as proposed in suppressing the single-scatter QCD background.

To study the dependence of the $\mpt$ on the instantanous luminosity, we choose as a proxy the number of primary collision vertices $N_{vtx}$ as reconstructed in the spectrometer. While this constitutes an approximation, since for any vertex reconstruction algorithm efficiency drops progressively as the luminosity increases, it is sufficient to understand the qualitative features of luminosity dependence.

\begin{figure}
    \includegraphics[width=7cm]{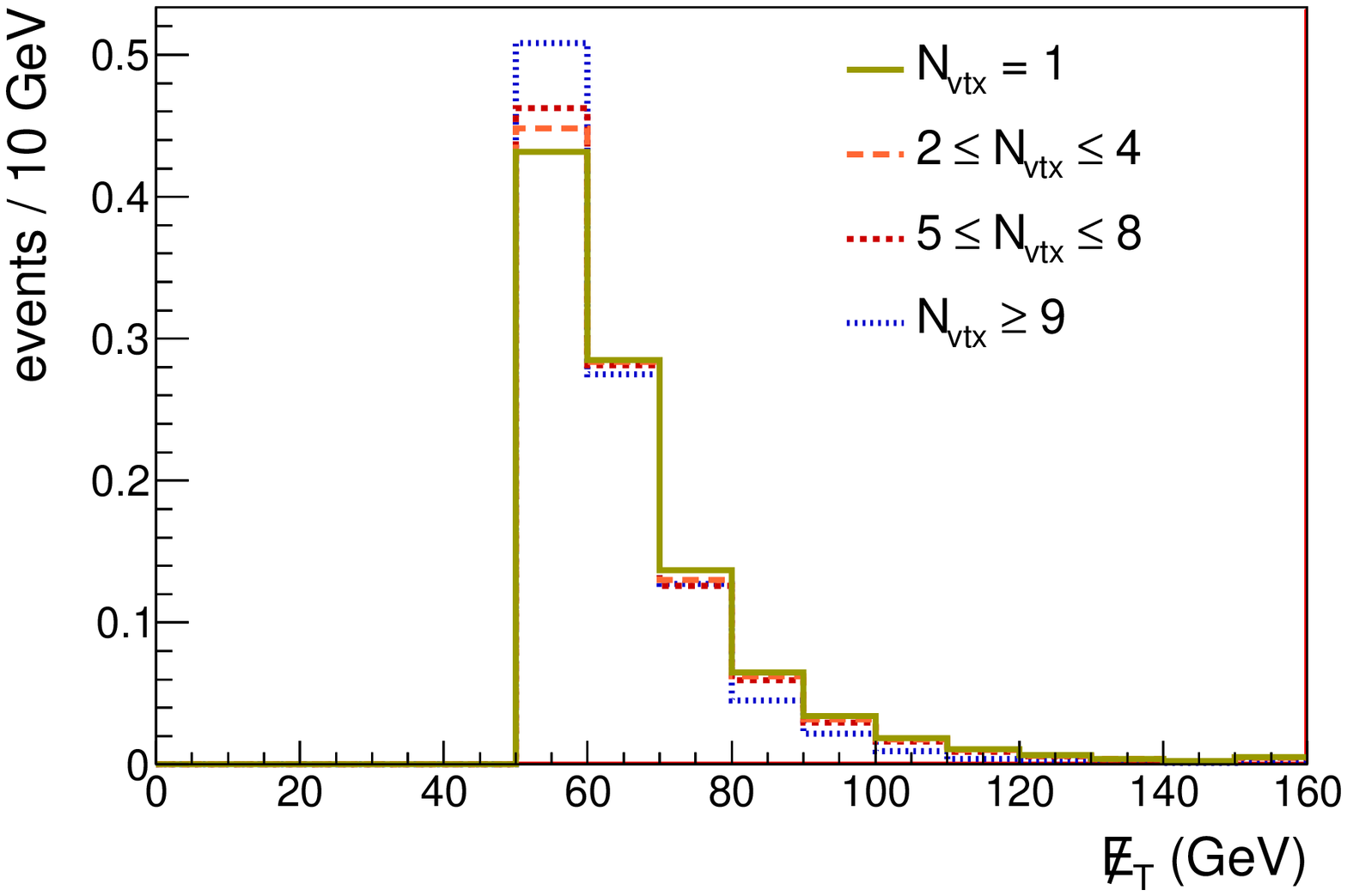}
    \includegraphics[width=7cm]{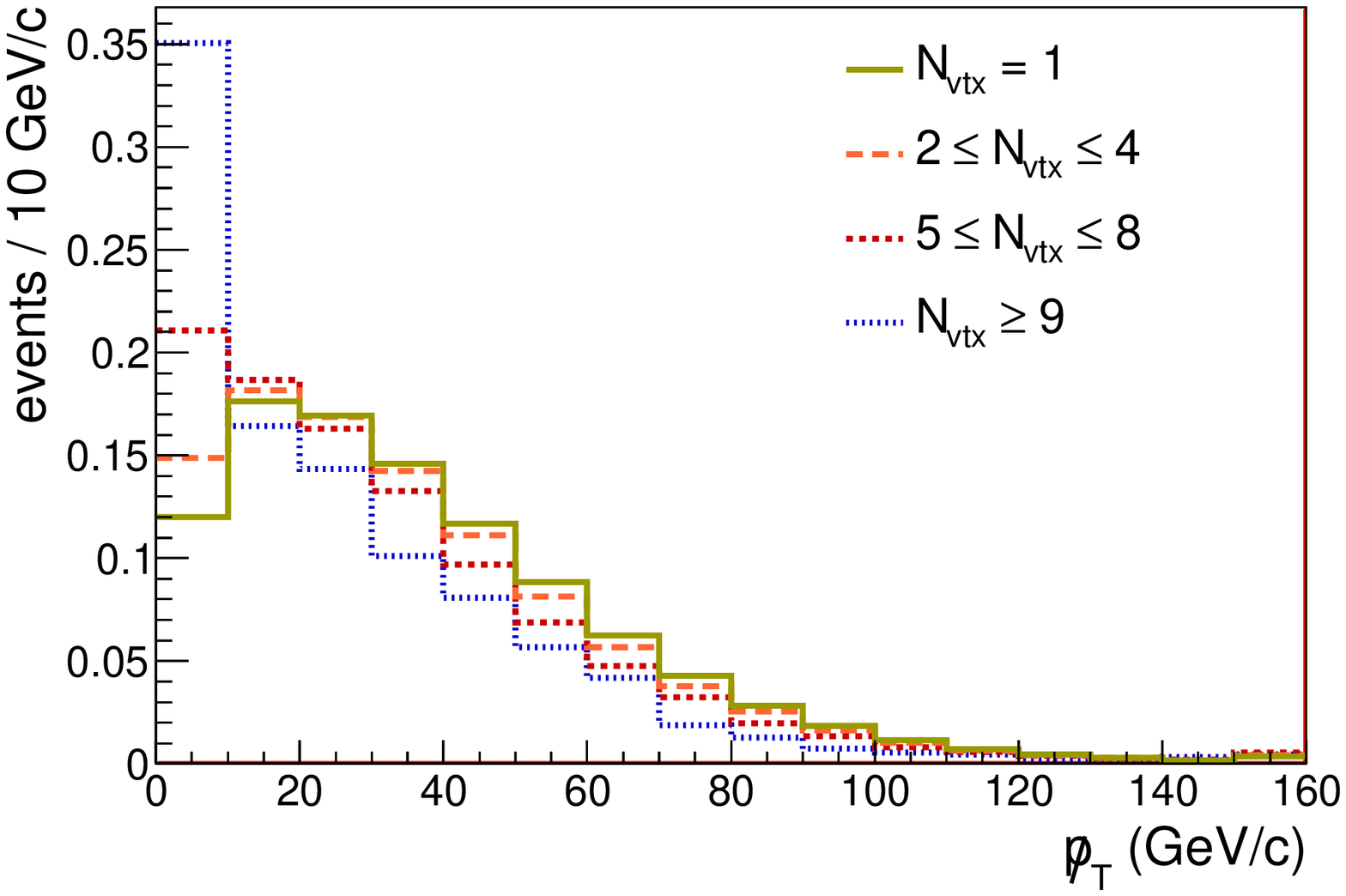}
    \includegraphics[width=7cm]{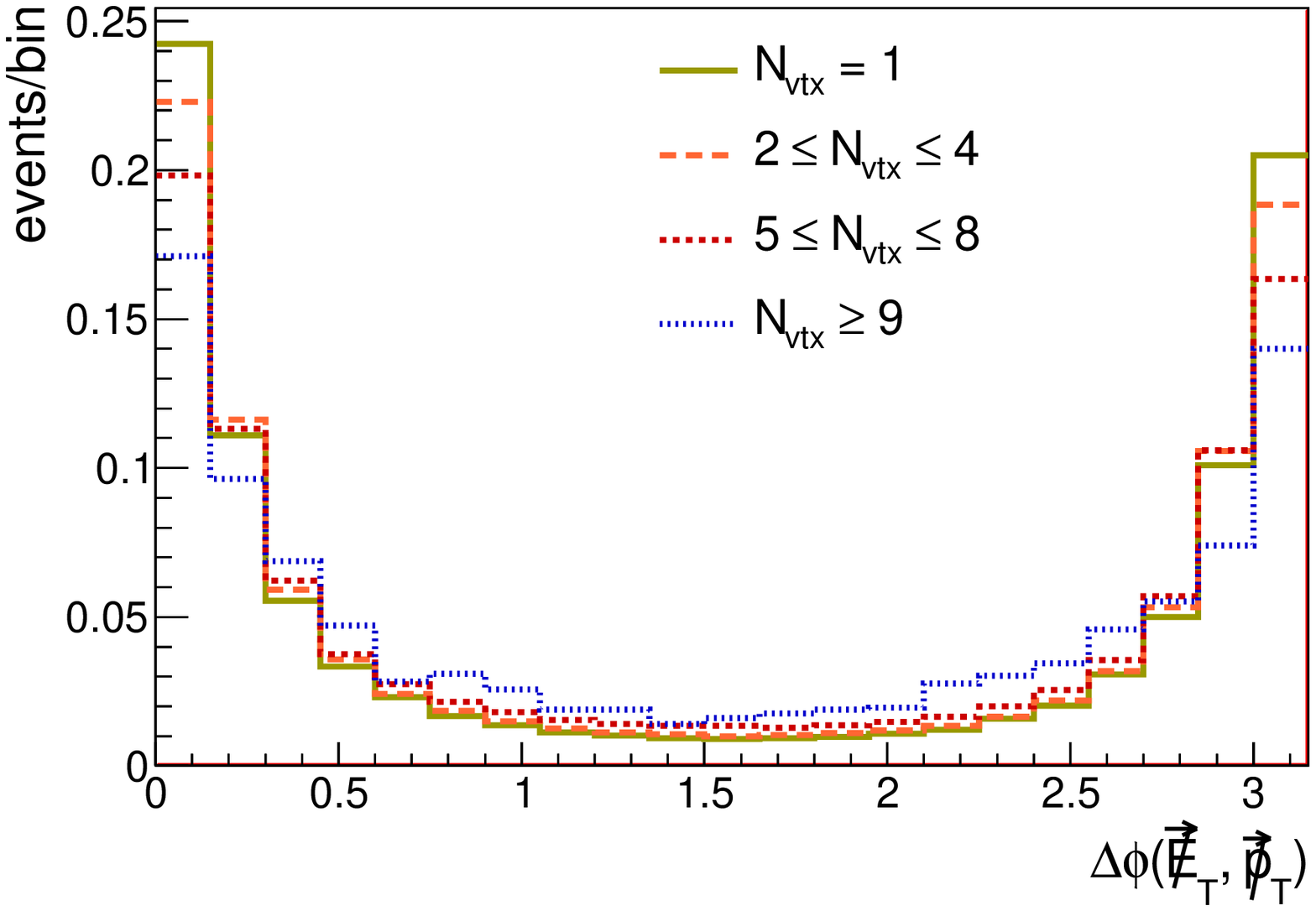}
    \caption{\label{fig:Nvtx} From top to bottom: the $\met$, $\mpt$ and $\Delta \phi(\vecmet, \vecmpt)$ distributions in events with varying number of interaction vertices. It is clear how the latter two plots show far better discrimination power toward the isolation of pile-up events with respect to the traditional cut on missing transverse energy, and can thus be used to control more efficiently the pile-up background. All plots are normalized to unit area.}
\end{figure}

Analyzing the data for different numbers of interaction vertices is possible to appreciate the pile-up rejection power of a cut on $\mpt$. This can be particularly interesting in a LHC context, where the far higher instantaneos luminosity makes extremely important to control this source of background.
Fig. \ref{fig:Nvtx} shows the $\met$, $\mpt$ and $\Delta \phi(\vecmet, \vecmpt)$ distributions dependence from the number of vertices for the whole data sample ({\it i. e.} for any number of jets). 
As can be seen, after fixing a certain online cut on $\met$, the offline $\met$ has only a mild dependence on the number of primary vertices. On the other hand, the pile-up events tend clearly to populate the region with QCD-like characteristics,  {\it i.e.} at low-$\mpt$ and simmetrically in the $\Delta \phi(\vecmet, \vecmpt)$ distribution.

\section{A $\Delta \phi(\vecmet, \vecmpt)$-based QCD data-driven model}

We stressed that the peculiar way QCD populates the $\Delta \phi(\vecmet, \vecmpt)$ distribution is by itself a powerful instrument to reduce the QCD background by a factor of two when cut at $\pi$/2, with minimal signal loss. Its implementation in more complex methods, like multivariate analysis, would maximize its impact. 

We argued, assisted by a MC simulation, that in the simplest case QCD populates the region where $\mpt$ is correlated or anti-correlated with $\met$ depending on the fraction of energy of the jets carried by charged particles. The fluctuations in the vectorial sum of charged particles' momenta in a jet, and in the amount of transverse momentum carried by those, derives from fluctuations in the parton shower and hadronization non-perturbative process. On the other hand, the $\met$ in QCD events appears preminently as a consequence of the fluctuation in the jet energy measurement in the calorimeter, or in the peculiar topology where one or more of the jets in the event falls in an uninstrumented region of the detector. The reasons for the appearance of the $\met$ and of the $\mpt$ in QCD events are completely uncorrelated. An angular correlation is on the other hand introduced by the fact that in the simple event topology of QCD dijet events, $\met$ and $\mpt$ have to be aligned with either one jet or the other. As demonstrated in a previous section, this topology is only mildly altered in the circumstance where QCD events appear with larger jet multiplicities due to radiation. 

An important consequence of the peculiar angular correlation between  $\met$ and $\mpt$ is that QCD is supposed to populate symmetrically the  $\Delta \phi(\vecmet, \vecmpt)$ distribution, with the symmetry centered at $\pi/2$. By rejecting the events with $\Delta \phi(\vecmet, \vecmpt) > \pi/2$ and defining as signal region the one with $\Delta \phi(\vecmet, \vecmpt) < \pi/2$, one can infer the contribution of QCD in the signal region in a data-driven fashion by counting the events populating the rejected region, i.e.  $\Delta \phi(\vecmet, \vecmpt) > \pi/2$. The robustness of this method has been tested by varying the jet multplicity, by placing cuts on the angular correlation among $\met$ and the jets direction, varying $\met$ cuts, etc. \cite{Marco_Tesi}. In all instances, by using MC simulation the above assumption has been found true to a level of $\sim 20\%$. The exact value should be determined by properly defining a ``control region" for the analysis. 

Following the same reasoning, it is legitimate to assume that QCD events in the $\Delta \phi(\vecmet, \vecmpt) \, > \, \pi/2$ could be used to model the kinematic features of QCD events in the complementary region.
The systematic uncertainties associated with the simulation of high multiplicity QCD jet production are very large. For these reasons a reliable data-driven modeling method for this category of events is of fundamental importance in collider physics analysis. 
This has been done in \cite{Abazov:2009jf, Aaltonen:2009fd} for dijet signatures and in \cite{Aaltonen:2011na, Marco_Tesi} with multiplicities up to 10. 
In all instances, a general good agreement has been found for a large number of the kinematical and topological distributions.

\section{Outlook on the usage of $\mpt$ at the ATLAS and CMS detectors}

The ATLAS and CMS calorimeter performance are comparable or mildly superior to the one of the CDF detector used in this study. This means that QCD multijet production is still a serious obstacle to physics measurement and searches in the $\met+jets$ signature. On the other hand, the coverage of the ATLAS and CMS charged particle spectrometer is far superior to the CDF one. Both arguments lead the authors to believe that the application of the techniques presented here to these experiments would be straightforward, and bring notable advantages. In particular, the authors call for a revisitation of the current ATLAS and CMS strategies of tightening the $\met$ requirements to cope with the increasing collider instantaneous luminosity, and suggest and alternative way to control the QCD background. The authors are aware that CMS uses already the spectrometer in a ``particle flow" reconstruction in order to increase the $\met$ and jet energy resolution; still, improving the resolution on those observables reduces the QCD contribution to $\met+jets$ samples but does not fully eliminate it. We argue that even after the use of particle flow techniques the $\mpt$ and its correlation with $\met$ will help to suppress QCD and to model its contribution. 
It is foreseeable that future upgrades of the ATLAS and CMS trigger systems could allow the computation of a $\mpt$ ``primitive" online, in order to contain the $\met$ and $\met+jets$ trigger rates to reasonable levels, without having to raise $\met$ cut to the point where the loss of interesting signals become unacceptable.

\section{Conclusions}
We introduced here a new observable that can be used to infer informations on the four-momentum of invisible particles passing trough collider detectors. 
The paper focuses on collisions where invisible particles appear together with quarks or gluons in the final state, where improvements in the identification techniques of invisible particles is crucial in order to suppress the otherwise dominant QCD background. 

As the technique uses only the spectrometer, as opposed to the traditional use of the calorimeter, the instrumental resolutions are completely uncorrelated. In particular, the resolution on the $\mpt$ measurement depends almost solely on the inherent fluctuation on the number and transverse momentum carried by the charged hadrons of the jets in the event. 

We have further shown the importance of the evaluation of the angular correlation between $\vecmpt$ and $\vecmet$, as a valid tool to reduce the contribution of the QCD multijet background in $\met$+jets signatures, studying the properties of $\mpt$-related observables in collider data and in MC simulations. By taking advantage of the peculiar properties of $\mpt$, we developed a data-driven way to understand the QCD contribution to the $\met+jets$ sample with relatively large precision, and suggest that in a similar fashion one can reproduce with similar level of detail the kinematic properties of QCD multijet events. The techniques developed here have been already successfully used in a broad range of analyses at the Tevatron experiments\,\cite{CDF_nunubb, Abazov:2009jf, Aaltonen:2011na, Aaltonen:2009fd, Aaltonen:2009jj, Aaltonen:2010fs,  Aaltonen:2012tq, Abazov:2010wq}. We argue that their use will be especially important in the searches for new phenomena at the LHC. 

\section{Acknowledgements}

We thank Marco Rescigno for the useful discussion. We also thank the CDF collaboration, the Fermilab staff and the technical staffs of the participating institutions for their vital contributions. This work was supported in part by the US Department of Energy. MB is supported by a fellowship of the Fondazione Angelo Della Riccia. The research of FM is supported in part by the Ministero Istruzione Universit\`a e Ricerca and the Istituto Nazionale di Fisica Nucleare.

\bibliographystyle{model1a-num-names}

\begin{thebibliography}{00}




\bibitem{CDF_nunubb}
  T.~Aaltonen {\it et al.}  [CDF Collaboration],
  Phys.\ Rev.\ Lett.\  {\bf 104}, 141801 (2010)
  arXiv:0911.3935.

\bibitem{Abazov:2009jf}
  V.~M.~Abazov {\it et al.}  [D0 Collaboration],
  Phys.\ Rev.\ Lett.\  {\bf 104} (2010) 071801
  arXiv:0912.5285.



\bibitem{Fox:2011pm} 
  P.~J.~Fox, R.~Harnik, J.~Kopp and Y.~Tsai,
  Phys.\ Rev.\ D {\bf 85}, 056011 (2012)
  arXiv:1109.4398.

\bibitem{Aaltonen:2012jb}
  T.~Aaltonen {\it et al.}  [CDF Collaboration],
  arXiv:1203.0742.

\bibitem{Chatrchyan:2011nd}
  S.~Chatrchyan {\it et al.}  [CMS Collaboration],
  Phys.\ Rev.\ Lett.\  {\bf 107} (2011) 201804
  arXiv:1106.4775.



\bibitem{Papucci} 
  M.~Papucci, J.~T.~Ruderman and A.~Weiler,
  arXiv:1110.6926.


\bibitem{ATLAS_sbottom}
  G.~Aad {\it et al.}  [ATLAS Collaboration],
  arXiv:1112.3832.


\bibitem{CDF}
  D.~Acosta {\it et al.}  [CDF Collaboration],
  Phys.\ Rev.\ D {\bf 71} (2005) 032001.

\bibitem{COT}
  T. Affolder {\it et al.},
  Nucl. Instrum. Methods A {\bf 526}, 249 (2004).

\bibitem{emcal}
L. Balka et al. (CDF Collaboration), Nucl. Instrum.
Methods A 267, 272 (1988); M. G. Albrow et al.
(CDF Collaboration), Nucl. Instrum. Methods A 480,
524 (2002).

\bibitem{hadcal}
S. Bertolucci et al. (CDF Collaboration), Nucl. Instrum.
Methods A 267, 301 (1988).

\bibitem{jetclu}
  F. Abe {\it et al.} [CDf Collaboration],
  Phys. Rev. D {\bf 45}, 001448 (1992).

\bibitem{jetcorr}
  A. Bhatti {\it et al.},
  Nucl. Instrum. Methods A {\bf 566}, 375 (2006).

\bibitem{Aaltonen:2011na}
  T.~Aaltonen {\it et al.}  [CDF Collaboration],
  Phys.\ Rev.\ Lett.\  {\bf 107} (2011) 191803.




\bibitem{alpgen}
  M. Mangano et al., J. High Energy Phys. 0307 (2003)
  001.

\bibitem{pythia} T. Sjostrand {\it et al.}, Comput. Phys. Commun. {\bf
    238}  135 (2001), version 6.422.


\bibitem{matching_scheme}
M. L. Mangano, M. Moretti, F. Piccinini and
M. Treccani, J. High Energy Phys. 0701 (2007) 013.


\bibitem{WZ_crosssections}
J.~M.~Campbell and R.~K.~Ellis, Phys.\ Rev.\ D {\bf 60}, 113006 (1999).

\bibitem{Lancaster:2011wr} 
  Tevatron Electroweak Working Group and for the CDF and D0 Collaborations,
  arXiv:1107.5255.
 
\bibitem{Galtieri:2011yd} 
  A.~B.~Galtieri, F.~Margaroli and I.~Volobouev,
  Rep. Prog. Phys. 75 (2012) 056201.


\bibitem{ttbar_cross_section}
  S.~Moch and P.~Uwer,
  Nucl.\ Phys.\ Proc.\ Suppl.\  {\bf 183} (2008) 75.




\bibitem{GEANT}
R. Brun {\it et al.} GEANT 3 manual, CERN Program Library Long Writeup. 1994.


\bibitem{Marco_Tesi}
  M.~Bentivegna,
  FERMILAB-MASTERS-2011-04.


\bibitem{Aaltonen:2009fd}
  T.~Aaltonen {\it et al.}  [CDF Collaboration],
  Phys.\ Rev.\ Lett.\  {\bf 103}, 091803 (2009).





\bibitem{Aaltonen:2009jj}
  T.~Aaltonen {\it et al.}  [CDF Collaboration],
  Phys.\ Rev.\ Lett.\  {\bf 103} (2009) 092002.

\bibitem{Aaltonen:2010fs}
  T.~Aaltonen {\it et al.}  [The CDF Collaboration],
  Phys.\ Rev.\  D {\bf 81} (2010) 072003.



\bibitem{Aaltonen:2012tq} 
  T.~Aaltonen {\it et al.}  [CDF Collaboration],
  arXiv:1203.4171.

\bibitem{Abazov:2010wq}
  V.~M.~Abazov {\it et al.}  [D0 Collaboration],
  Phys.\ Lett.\ B {\bf 693}, 95 (2010).


 \end{thebibliography}

\end{document}